\documentclass[11pt,a4paper]{article}
\usepackage[utf8]{inputenc}
\usepackage{amsmath}
\usepackage{amsfonts}
\usepackage{amssymb}
\usepackage{graphicx}
\usepackage{dsfont}
\usepackage{amssymb}
\usepackage[margin=1.0in]{geometry}
\usepackage{algorithm}
\usepackage{algpseudocode}
\usepackage{ upgreek }
\usepackage{pifont}
\usepackage{url}
\usepackage{titling}
\usepackage{authblk}
\usepackage[sort&compress,numbers,square]{natbib} 
\newcommand{\code}[1]{\texttt{#1}}
\usepackage[table,xcdraw]{xcolor}
\usepackage{multirow}
\usepackage[normalem]{ulem}
\newcommand\redout{\bgroup\markoverwith {\textcolor{red}{\rule[0.5ex]{2pt}{0.8pt}}}\ULon}

\graphicspath{ {./figures/} }

\DeclareMathOperator*{\argmin}{arg\,min}
\renewcommand{\d}{\mathrm{d}}

\usepackage[colorinlistoftodos]{todonotes}

\usepackage{float}

\usepackage{verbatim}

\usepackage{soul} 

\usepackage{tikz}
\newcommand{\tikznode}[2]{%
\ifmmode%
\tikz[remember picture,baseline=(#1.base),inner sep=0pt] \node (#1) {$#2$};%
\else
\tikz[remember picture,baseline=(#1.base),inner sep=0pt] \node (#1) {#2};%
\fi}
\newcommand{\beginsupplement}{%
        \setcounter{table}{0}
        \renewcommand{\thetable}{S\arabic{table}}%
        \setcounter{figure}{0}
        \renewcommand{\thefigure}{S\arabic{figure}}%
        \setcounter{section}{1}
        \renewcommand{\thesection}{S\arabic{section}}%
        \setcounter{equation}{0}
        \renewcommand{\theequation}{S\arabic{equation}}%
     }

\makeatletter
\newcommand{\settitle}{\@maketitle}
\makeatother

\begin{document}

\title{Unlabelled landmark matching via Bayesian data selection, and application to cell matching across imaging modalities}
\author[1, 2]{Jessica E. Forsyth}
\author[2]{Ali H. Al-Anbaki}
\author[2]{Berenika Plusa}
\author[1]{Simon L. Cotter\thanks{simon.cotter@manchester.ac.uk}}
\affil[1]{Department of Mathematics, University of Manchester, Manchester M13 9PL, United Kingdom}
\affil[2]{Faculty of Biology Medicine and Health, University of Manchester, Manchester M13 9PL, United Kingdom}
\date{}
\setcounter{Maxaffil}{0}
\renewcommand\Affilfont{\itshape\small}

\sffamily
\maketitle

\begin{abstract}
We consider the problem of landmark matching between two unlabelled point sets, in particular where the number of points in each cloud may differ, and where points in each cloud may not have a corresponding match. We invoke a Bayesian framework to identify the transformation of coordinates that maps one cloud to the other, alongside correspondence of the points. This problem necessitates a novel methodology for \emph{Bayesian data selection}; simultaneous inference of model parameters, and selection of the data which leads to the best fit of the model to the majority of the data. We apply this to a problem in developmental biology where the landmarks correspond to segmented cell centres, where potential death or division of cells can lead to discrepancies between the point-sets from each image. We validate the efficacy of our approach using \textit{in silico} tests and a microinjected fluorescent marker experiment. Subsequently we apply our approach to the matching of cells between real time imaging and immunostaining experiments, facilitating the combination of single-cell data between imaging modalities. Furthermore our approach to Bayesian data selection is  broadly applicable across data science, and has the potential to change the way we think about fitting models to data.
\end{abstract}






\section{Introduction}

Understanding early mammalian development is key to the advancement of in vitro fertilisation (IVF) techniques and improved understanding of early cell specification within mammals. Within developmental biology there have been significant advances in experimental techniques, including the ability to culture preimplantation embryos \textit{ex vivo} and monitor their development through periodic 3D imaging, known as real time imaging (RTI)~\cite{Abe2013, grabarek2012}. 
In conjunction with the generation of mouse reporter lines, such as the H2b:GFP line, we are able to visualise the development of the murine embryo and monitor the behaviour of individual cells~\cite{Hadjantonakis2004}. 
One of the highly disputed questions regarding the development of the preimplantation embryo, is the effect of cell history and changes in embryo architecture on cell lineage specification~\cite{plusa2020,fischer2020, Forsyth2021}.

After RTI experiments, embryos can be fixed to halt their development and stained for proteins of interest via immunostaining. The cells' respective fates can then be inferred from their protein expression profiles. In order to interrogate the relationship between cell history and cell specification it is crucial to link historical information (gained from RTI experiments) with protein expression (quantified from immunostaining) at the single cell level. However, the cell-to-cell matching across these two imaging modalities is non-trivial due to the random re-orientation of the embryo during staining and potential deformation during the fixation process. Coordinates of cell centres can be extracted from the final frame of the RTI experiment, using the H2b:GFP signal, and from the immunostained image, using the nuclear stain, allowing us to frame this problem as an unlabelled landmark matching problem where the landmarks are cell positions in three dimensions. 

 
It is most common for landmark-based registration to be approached using variational techniques~\cite{GutierrezBecker2017, kent2004matching}, although some probabilistic approaches have also been developed~\cite{dryden2007statistical, green2006bayesian}. A commonly used approach to match landmarks subject to some non-rigid transformation is the large deformation diffeomorphic metric matching (LDDMM) method. This method uses a curve to describe diffeomorphic mapping of individual landmarks between the target and template point clouds~\cite{Younes2009, Joshi2000}. A Bayesian approach of shape matching via a non-linear deformation is presented in \cite{Cotter2013}, where the geodesic map which takes one shape to the other is inferred. 

In this work we invoke the Bayesian framework in order to be able to not only find likely cell matchings, but also to quantify the uncertainty in those matchings. Our biological example has an additional difficulties, since the landmarks are unlabelled,  and the assumption that all landmarks exist in both point-sets does not hold. This discrepancy in landmarks can occur due to cell death or division between the time that the RTI experiment was stopped and fixation, or due to inaccurate segmentation of cell centres. One approach would be to manually clean the data and select only cells with guaranteed matches in the corresponding image, however this is highly subjective with potential for significant errors as we do not know \textit{a priori} which cells to eliminate from the registration. 

There has been some work on data selection with regards to single and multi-source data acquisition~\cite{Rahm2000}, however there is little Bayesian description of data selection, although there are similarities with Bayesian model selection~\cite{ando2010bayesian} and outlier detection~\cite{aggarwal2017introduction}. In this work, we introduce a novel approach to \emph{Bayesian data selection}, which limits the effect that cells which do not appear in both images have on the inference. This is implemented through the introduction of  parameters which describe our belief in the fidelity of each observation in the data. The values of these fidelity parameters are jointly inferred alongside the model parameters describing the transformation and correspondence of the landmarks.

We implement Markov chain Monte Carlo (MCMC) methods to explore the complex distribution on the model and fidelity parameters. The posterior is frequently highly multimodal, preventing complete exploration of the parameter state space due to `trapping' in local minima. We therefore implement tempering of the likelihood, alongside adaptive MCMC~\cite{roberts2009examples} to optimise our sampling and minimise trapping. 

Although the introduction of data selection is primarily introduced to facilitate landmark registration within our specific biological example, it is clear that this framework could be expanded to a very broad range of inferential problems, with potential for wide-ranging impact in many applications of data science.

In Section~\ref{sec:cell_matching} we introduce the transformation model, including the description of a 3D affine transformation and a non-linear deformation, and a method of describing landmark correspondence within the model. In Section~\ref{sec:bayesian_setup} we describe the construction of the posterior distribution that we wish to characterise. In Section~\ref{sec:data_selection} we introduce the concept of \emph{Bayesian data selection} and its implementation within the landmark registration problem. We then go on to describe the MCMC implementation in Section~\ref{sec:mcmc scheme}. In Section~\ref{sec:results} we firstly present several \textit{in silico} test problems demonstrating the efficacy of the approach. We then perform inference on embryos with microinjected fluorescent
cells which enable us to identify a subset of the cells in both images for validation on a real data set. We then demonstrate the accuracy of our approach on a problem in which we wish to match cells from the final frame of an RTI experiment with corresponding immunostained images, with the additional challenge of embryo matching. We conclude with a discussion in Section~\ref{sec:discus}.

\section{Landmark matching}\label{sec:cell_matching}
\begin{figure}[htpb]
\centering
\includegraphics[width=1.0\textwidth]{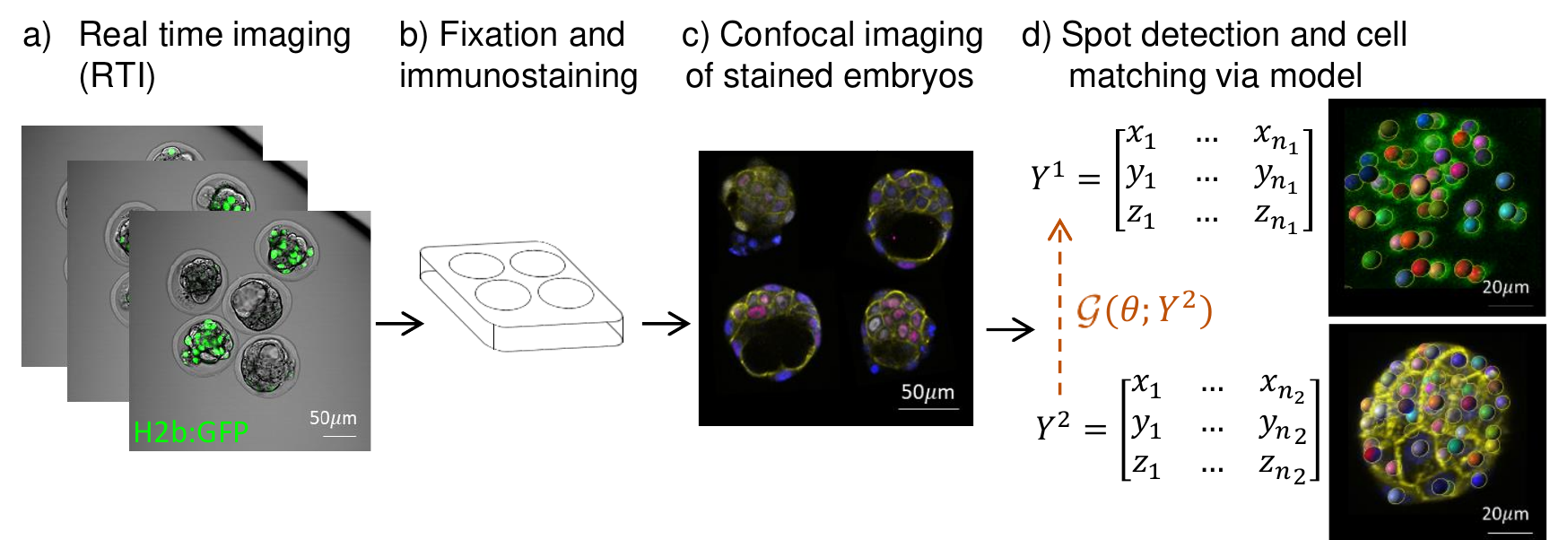}
\caption{Experimental design with examples of two-dimensional slices from images. Spot detection using IMARIS (BitPlane) from final frames of RTI (using H2b:GFP signal) and immunostained image (Hoechst). Cell matching via observation operator $\mathcal{G}(\theta;Y^2)$. Experimental protocols for embryo collection, visualisation and staining can be found in Sections (\ref{supp:embryocoll} -~\ref{supp:rti}).}
\label{fig:methodology}
\end{figure}

In order to better understand mammalian development, spatiotemporal information from RTI experiments must be linked with protein expression which is inferred from secondary immunostaining images. In order to link this data, cell centres are extracted from the final frame of the RTI study and the immunostained image and matched, Figure~\ref{fig:methodology}. Previously, attempts have been made to manually match the cells between images, however this is non-trivial due to the manipulation of the embryos during staining and can lead to low confidence matchings of cells between images. 

We can generalise this problem to the matching of two unlabelled point clouds:
\begin{equation}
Y^1 \in \mathds{R}^{d \times n_1}, \qquad
Y^2 \in \mathds{R}^{d \times n_2},
\end{equation}
 where $d \in \mathbb{N}$ is the dimension of the observation space, and the number of points in $Y^1$ and $Y^2$ is $n_1$, $n_2$ respectively where we assume $n_1\leq n_2$. In the context of cell matching, potential differences in $n_1$ and $n_2$ may arise from cell death or division after the completion of the RTI prior to fixation of the embryos, or due to segmentation errors. $Y^1$ and $Y^2$ are pre-processed such that the average coordinate in each cloud is shifted to $(0,0,0)$, and re-scaled through division by the minimum cell-to-cell Euclidean distance.
 
The two point clouds can be considered to be noisily transformed versions of each other, with labels subject to a random permutation, along with the potential addition or subtraction of points in both clouds. The transformation of $Y^2$ to $Y^1$ can be described by the composition of an affine transformation, a non-linear deformation, and a permutation of labels, described by an observation operator $\mathcal{G}(\theta;Y^2)$ with parameters $\theta$.

\subsection{Non-linear deformations via geodesic motion}\label{sec:localdef} 
Deformation can occur due to continued growth of the embryo prior to fixation, or manipulation of the embryo during immunostaining, necessitating an explicit description of a non-linear deformation within $\mathcal{G}$. The non-linear deformation to the point-set is modelled as a geodesic transformation resulting from the application of an initial momentum, $\mathbf{p}_i \in \mathbb{R}^d$ to each point $\mathbf{q}_i \in \mathbb{R}^d$~\cite{Bock2020,younes2019}. The deformed points, $\phi(\theta;Y^2) = \mathbf{q}(1)$, are then obtained through the ODE given by:
\begin{subequations}
  \begin{align} 
         \frac{d p_t^j}{d t}  &= \left(-\sum_{i=1}^M \frac{(q_t^i-q_t^j)}{\sigma_K^2} \exp\left( -\frac{\| q_t^i - q_t^j \|_2^2}{2 \sigma_K^2} \right) p_t^i \right)^\top \cdot p_t^j\label{eq:deftosolve_pos},\\
         \frac{d q_t^j}{dt} &= \sum_{i=1}^M \exp\left( -\frac{\| q_t^i - q_t^j \|_2^2}{2 \sigma_K^2} \right) p_t^i ,  \label{eq:deftosolve_mom}
\end{align}
\end{subequations} 
over the time interval $t=[0,1]$, where $\mathbf{q}(0)=Y^2$, details given in Section~\ref{supp:deformation}.

The application of the geodesic flow is computationally expensive due to the solving of $3n$ differential equations. We envisage that for smaller embryos, deformation is minimal, in which case we set $\phi$ to be the identity. However for larger embryos, it may not be possible to accurately match cells without the addition of inference of a non-linear transformation between $Y^1$ and $Y^2$.

\subsection{Affine transformation}
$\mathcal{G}$ also incorporates a three dimensional affine transformation to account for shear scaling, rotation and translation of points. The transformation is applied to $\phi(\theta;Y^2)$ to give
\begin{equation}
   \mathcal{F} (\theta,Y^2)=A(\theta)\phi(\theta;Y^2) + b(\theta)\mathbf{1}_{n_2}^\top,
\end{equation}
where $\mathbf{1}_{n_2} \in \mathbb{R}^{n_2}$ is a column vector of ones, $A(\theta)$ is the transformation matrix and $\mathbf{b}(\theta)$ the translation vector, see Section \ref{supp:affine}.

\subsection{Permutation of labels}
 Our overall aim is to find the labelling of points in order to match cells across images. We introduce a permutation vector as a method of describing the matching of cells across $Y^1$ and $Y^2$. The permutation vector $P\in \mathbb{N}^{n_2}$ 
 describes the ordering of cells in $Y^2$ in order to match them with cells in $Y^1$. Note that in the case that $n_1<n_2$, the cell numbers in the $n_2-n_1$ last entries of the permutation vector are assumed not to have a corresponding match in $Y^1$, and as such are not required for the calculation of the likelihood.

Our aim is to compare the positions of points in $Y^1$ with their corresponding matches, as given by $P$, in the transformed cell centres in $Y^2$. As such, we define the matrix $M_P \in \mathbb{R}^{n_2 \times n_1}$ where
\[
M_P = \begin{pmatrix} e_{P_1} & e_{P_2} & \ldots & e_{P_{n_1}} \end{pmatrix},
\]
where $e_i \in \mathbb{R}^{n_2 \times 1}$ are the standard canonical basis column vectors for $\mathbb{R}^{n_2}$. Post multiplication of the transformed $Y^2$ coordinates by $M_P$ gives us a matrix of the new cell center positions ordered according to $P$.

\subsection{The observation operator}
We define our observation operator $\mathcal{G}:\Theta \to \mathbb{R}^{3 \times n_1}$, which takes the cell center coordinates of $Y^2$, applies a non-linear transformation (if being applied), an affine transformation, and then reorders the subset of the cells which we aim to match to $Y^1$ according to the permutation vector $P$. Therefore we arrive at

\begin{equation}
 \mathcal{G}(\theta;Y^2) = \left ( A(\theta)\phi(\theta;Y^2) + b(\theta) \mathbf{1}_{n_2}^\top \right )M_P(\theta).
 \end{equation}

\section{Bayesian cell matching} \label{sec:bayesian_setup}
Bayes' theorem is a fundamental property of sets and measures that forms the basis of a probabilistic framework for inverse problems, involving the combination of prior knowledge, observations, and models. Our aim is to characterise the posterior probability density $\pi(\theta|Y)$ on the model parameters $\theta$ conditioned on the data $Y$, which by Bayes' theorem is given by:
\begin{equation}
    \pi(\theta|Y) \propto L(Y|\theta) \pi_0(\theta),
\end{equation}
where $L(Y|\theta)$ is the likelihood of the observations given $\theta$, and $\pi_0(\theta)$ is the prior density.

The matching of cells between images can be considered as an inverse problem where we wish to identify a transformation of $Y^2$ in order to identify the correct matching, $P$, of the cells. The inverse problem of cell matching is complex with potentially correlated parameters across the components of the model, leading to potentially multimodal posterior distributions.  

\subsection{The likelihood}
We assume that the observations of the cell centers are subject to mean-zero i.i.d. Gaussian noise, such that:
\begin{equation}
Y^1_i=\mathcal{G}_i(\mathbf{\theta};Y^2) + \eta_i, \qquad \eta_i \sim \mathcal{N}(0,\Sigma),
\end{equation}
where $Y^1_i$ is the $i$th column of $Y^1$, and $\mathcal{G}_i(\mathbf{\theta})$ is the $i$th column of $\mathcal{G}(\theta)$. Therefore the likelihood is given by:
\begin{equation}
L(Y^1, Y^2|\mathbf{\theta},\Sigma) \propto \prod_{i=1}^{n_1} \exp \left( -\frac{1}{2} \left\| Y^1_i  - \mathcal{G}_i(\theta;Y^2)\right\|_\Sigma ^2\right),
\label{eq:likelihood}
\end{equation}
where $\|x\|^2_\Sigma = x^\top \Sigma^{-1} x$ is the covariance-weighted norm.

\subsection{Priors}

We choose mean zero priors on the affine transformation parameters and deformation momenta as shown in Table~\ref{tab:prior}, and a uniform prior on label permutations. 
\begin{table}[htpb]
\centering
\begin{tabular}{|c|c|c|}
\hline
\textbf{Parameter} & \textbf{Prior }      & \textbf{ Parameters}                                                  \\ \hline
$\alpha_i, \beta_j$         & $\mathcal{U}(-\pi,\pi)$          &                                                                          - \\ \hline
$s_i$              & $\mathcal{N}(0,\sigma_s)$        & $\sigma_s=0.1$                                                             \\ \hline
$b_i$              & $\mathcal{N}(0,\sigma_b)$        & $\sigma_b=0.1$                                                               \\ \hline
$p_i$              & $\mathcal{N}(0,\sigma_p)$        & $\sigma_p=1$                                                               \\ \hline
$P$              & $\mathcal{U}(S_{n_2})$        & -                                                          \\ \hline
\end{tabular}
\caption{Prior distributions. $S_{n_2}$ denotes the symmetric group of all possible permutations of $\{1, \ldots, n_2\}$.}
\label{tab:prior}
\end{table}



\subsection{Hierarchical Bayes posterior}

The noise covariance $\Sigma$ is unknown \textit{a priori} and so we use a hierarchical Bayes approach to infer its value alongside the model parameters. We choose the Inverse-Wishart distribution as a prior on $\Sigma$ which is conjugate to the Gaussian likelihood, enabling marginalisation of $\Sigma$~\cite{alvarez2014, Liu2015}. This distribution has two parameters, the degrees of freedom $\nu>d-1$, and the positive definite symmetric scale matrix $\Psi \in \mathbb{R}^{d \times d}$. The Inverse-Wishart distribution has a mean given by
\begin{equation}
\mathds{E}(\Sigma)=\frac{\Psi}{\nu-d-1},
\end{equation}
when $\nu>d+1$, and variance of the diagonal terms given by
\begin{equation}
\mathrm{Var}(\Sigma_{ii}) =\frac{2\Psi_{ii}^2}{(\nu -d-1)^2(\nu -d-3)},
\end{equation}
when $\nu>d+3$.
We choose $\nu$ and $\Psi \propto I_3$ to achieve $\mathds{E}(\Sigma)= 0.01I_3$ and $\mathrm{Var}(\Sigma_{i,i})=0.2^2$, giving us $\nu=6.0050$ and $\Psi=0.0201 I_3 $. 

The posterior given by:
\begin{equation} 
\pi(\theta | Y^1, Y^2) \propto L(Y^1, Y^2 |\theta, \Sigma) \pi_0(\theta) \pi_0(\Sigma),
\label{eq:postpred}
\end{equation} 
can be marginalised with respect to $\Sigma$ to give the target density:
\begin{subequations}
\begin{align}
    \pi(\theta | Y^1, Y^2) &\propto \pi_0(\theta) \int_\Omega L(Y^1, Y^2 |\theta, \Sigma) \pi_0(\Sigma) d\Sigma,\\
     &\propto \pi_0(\theta) \int_\Omega \prod_{i=1}^{n_1} \exp \left( -\frac{1}{2} \left\| Y^1_i  - \mathcal{G}_i(\theta;Y^2)\right\|_\Sigma ^2\right) \mathcal{W}^{-1}(\Sigma) d\Sigma,\\
     &\propto \pi_0(\theta) \; \left| \Psi + (Y^1  - \mathcal{G}(\theta;Y^2)) (Y^1  - \mathcal{G}(\theta;Y^2)) ^\top \right| ^{\frac{-\nu + n_1}{2}}.
\end{align}
\end{subequations}

\section{Bayesian Data Selection} \label{sec:data_selection}

The observation operator $\mathcal{G}$ describes the transformation and permutation of points in $Y^2$ to match $Y^1$, but assumes that all cells in $Y^1$ have a corresponding match in $\mathcal{G}(\theta;Y^2)$. This assumption does not always hold, since cells can divide or undergo apoptosis in between the RTI experiment and fixation, or may be too faint for accurate segmentation, resulting in the presence of cells within one or both of the point clouds with no corresponding match. 
We cannot know which cells do not have a match \emph{a priori}, and therefore we aim to infer this information, thereby conducting what we will refer to as \emph{Bayesian data selection}. This refers to any approach where additional parameters are introduced into the inference which dictate the sensitivity of the posterior to a given observation, where the values of these parameters are themselves inferred from data, jointly with the model parameters.

\subsection{Data fidelity} 
We aim to infer the value of parameters that represent our belief in the  fidelity of each observation along with model parameters. We introduce fidelity parameters $\gamma_i \in [0,1]$ for each observation (in our case a cell center from $Y^1$),  that controls the relative contribution of that observation to the likelihood. These $\gamma_i$ are effectively inverse annealing temperatures for each observation, with high temperatures (where $\gamma_i\ll 1$) resulting in a likelihood which is not sensitive to the data-model mismatch for this observation. This approach limits the effect on the posterior of spurious data, for instance in the case where a cell has no corresponding match.

The likelihood function is ordinarily a function of the mismatch between each observation $Y_i$, and the observation operator at a given value of the model parameters $\mathcal{G}_i(\theta)$, such that
\begin{equation}\label{eq:OB}
    L(Y|\theta) = f_L(Y_1 - \mathcal{G}_1(\theta), \ldots , Y_n - \mathcal{G}_n(\theta)).
\end{equation}
In ordinary Bayesian inference the likelihood is sensitive to all of the data-model mismatches $Y_i - \mathcal{G}_i(\theta)$, which causes issues when the data is corrupted, or where the model does not adequately describe the entirety of the data. We now aim to infer which of these data can be well-matched to the model through a likelihood which takes into account the fidelity of each observation, given by:
\begin{equation}\label{eq:BDS}
    L_\gamma(Y|\theta, \gamma = [\gamma_1, \ldots, \gamma_n]) = f_L(\gamma_1(Y_1 - \mathcal{G}_1(\theta)), \ldots , \gamma_n (Y_n - \mathcal{G}_n (\theta))).
\end{equation}
In the context of Bayesian unlabelled landmark matching, for each point in $Y^1$ the fidelity parameter $\gamma_i \in [0,1]$  represents our belief that this cell has a match in $Y^2$. A value of $\gamma_i=0$ corresponds to a likelihood which is independent of the data-model mismatch of the $i^{th}$ observation, and $\gamma_i=1$ corresponds to a likelihood which is dependent on the $i^{th}$ cell's mismatch as in equation \eqref{eq:OB}.

The inclusion of the fidelity parameters works to prevent the fitting of a model to an entire set of points for which a subset may not be adequately described by that model. Without appropriate data selection in the landmark matching problem,  there are no guarantees that the transformation and permutation that leads to the lowest overall least squares fit corresponds with the correct matching.

\subsection{Application to cell matching}

We choose a beta prior $B(\alpha,\beta)$ on each $\gamma_i$,  with $\alpha=2$ and $\beta=2$ such that $\mathds{E}(\gamma_i)=0.5$ and $\mathrm{Var}(\gamma_i)=0.05$ and consider the data fidelity posterior distribution density which arises from multiplying each data-model mismatch by its fidelity parameter $\gamma_i$:
\begin{equation}
    \pi(\theta, \Gamma | Y^1, Y^2) \propto \pi_0(\theta) \pi_0(\mathbf{\gamma})  \frac{1}{Z(\Gamma)}  \left| \Psi + (X \Gamma) (X \Gamma)^\top \right| ^{\frac{-\nu + n_1}{2}},
    \label{eq:finalposterior}
\end{equation}
where $\Gamma = \mathrm{diag}(\gamma_1, \ldots, \gamma_{n_1}) \in \mathbb{R}^{n_1 \times n_1}$, and $X=Y^1-\mathcal{G}(\theta;Y^2)$. We note that
\begin{equation} \label{eq:fidconstantterm}
    Z(\Gamma) \propto \left( \prod_{i=1}^{n_1} \gamma_i^{d} \right),
\end{equation}
as shown in Section \ref{supp:datafidconstant}.


\section{MCMC methodology}\label{sec:mcmc scheme} 

The posterior distribution is a highly complex multimodal distribution on a high dimensional space, involving a mixture of continuous and discrete variables. In order to generate samples from the posterior distribution, we implement MCMC, a commonly used approach to sample from complex probability distributions. As our model and data selection approach is inherently modular (transformation, permutation and fidelity modules), we use a Metropolis-within-Gibbs approach~\cite{tierney1994}, which enables us to tune the proposal variances adaptively for each of the modules involving continuous random variables. 


\subsection{Multimodality and tempering} 

We assume that the parameter state space is dominated by a single best-fit mode, corresponding to the correct matching of points. 
However, the state-space is likely to be multimodal and difficult to sample from due to its complexity and the level of correlation between components of the model. To facilitate better exploration of the parameter space and avoid trapping in local minima, we implement likelihood-tempering, as described in~\cite{marinari1992}. During early iterations improved mixing is promoted through a high temperature $T$, with adapted Metropolis-Hastings formula given by:
\begin{equation}\label{eq:mh}
        \alpha= \min \left(1, \frac{\pi_0(v)}{\pi_0(u)} \exp\left(\frac{1}{T} (\Phi_u - \Phi_v)\right)  \frac{Q(u|v)}{Q(v|u)} \right),
\end{equation}
where $\Phi$ is the negative log of the posterior predictive, and $Q(\cdot,\cdot)$ is the proposal kernel. The temperature $T>0$ is gradually reduced via an exponential cooling schedule. Selection of the start temperature $T_0$ and the cooling rate of the system is crucial to the successful and efficient identification of the dominant mode,
details given in~\ref{supp:tempering}. 
Once $T=1$, the temperature is fixed and subsequent samples from the posterior recorded.

\subsection{Proposals on continuous random variables}
A subset of the parameters we wish to infer are defined on bounded state spaces. In order to facilitate efficient proposals on these parameters we transform them onto an unbounded domain. The proposals are then mapped back to the bounded parameters, and the adjustment to the proposal density included in the calculation of the acceptance ratio, details given in~\ref{supp:boundedRW}.

Random walk proposals are used for the continuous random variables, given by
\begin{equation}\label{eq:rw}
    v=u+\beta \xi, \; \; \; \; \xi \propto \mathcal{N}(0,C),
\end{equation}
where $\beta$ is the step-size of the proposed move, and tuned so that we achieved the optimum 23.4\% acceptance rate within each Gibbs module~\cite{gelman1997weak}. $\beta$ is also adjusted during temperature reductions by a factor $(1-\sqrt{t_c})$, where $t_c$ is the cooling rate of the tempering schedule, to account for changes in the target distribution density. $\beta$ is initialised with a value $\frac{2.38^2}{n_p}$ where $n_p$ is the number of parameters being sampled on within that Gibbs iteration. 

The proposal covariance $C$ is initialised as a diagonal matrix of the prior variances. However, to facilitate better sampling on $\mathbf{\theta}$ at $T=1$, we implement adaptive MCMC where we use the sample covariance matrix of accepted samples to construct efficient proposals on the model parameters, details given in~\ref{supp:adaptivemcmc}. 

\subsection{Proposals on the permutation vector} 

MCMC techniques are predominantly designed for continuous problems, rather than for  discrete problems such as permutation sampling~\cite{Zanella2019}. In order to explore different permutation vectors, we use a proposal distribution that is uniform on a set of permutations which are one switch of labels different from the current state. When at an initial permutation vector $P$ we propose the swapping of two cell labels $i\neq j$ to generate $P'$. This proposal is uninformed and symmetric about $P$, therefore giving the same acceptance probability as a standard random-walk on continuous random variables.

\subsection{Interpretation of results}\label{sec:interpretation}
To interpret the results of our sampling on the permutation vector, we record the number of times each cell in $Y^1$ matches with each cell in $Y^2$ during sampling at $T=1$. The number of matches is recorded using a matrix $M \in \mathbb{R}^{n_1 \times n_2}$. The matrix is then normalised so that the entries represent the proportion of samples in each matching, which can be visualised using probability heatmaps. 

In order to calculate the most likely matching (MLM) of the cells in $Y^1$, we solve the linear assignment problem
\begin{equation}
    P_{\mathrm{MLM}} = \argmin_{P\in S_{n_2}} \sum_{i=1}^{n_1} A(i,P_i),
\end{equation}
where $A_{i,j}=1-M_{i,j}$, using the \code{matchpairs} MATLAB function~\cite{Duff2001}. From this we can describe the MLM of a given chain, and then compare this to the ground truth permutation vector for the \textit{in silico} tests. In tests using real data where the true matching is unknown, this MLM would be representative of the inferred matching of points for subsequent analyses.

To assess the accuracy of the spatial matching of the points, we evaluate and stored thinned samples of the cell-to-match distances given by
\begin{equation}
    d_i=\|Y^1_i - \mathcal{G}_i(\theta,Y^2) \|_2,
\end{equation}
during sampling at $T=1$. These values are then used to evaluate the median and root-mean-squared-error (RMSE) of the cell-to-match distances for each chain, thereby giving an indication of the spatial quality of the matchings.  

To allow us to visualise the quality of the MLM, and compare fidelity parameters of cells, we also calculate the MAP estimates on the transformation and fidelity parameters, conditioned on the MLM, details given in~\ref{supp:mapest}.

\section{Results} \label{sec:results}
We first constructed several \textit{in silico} tests which were designed to mimic real cell matching problems. The \textit{in silico} test problems used real cell centre coordinates segmented from images of fixed embryos for $Y^2$. We chose to use embryos from four key stages within the mammalian preimplantation period with; 8, 15, 33 and 62 cells respectively,  see~\ref{supp:embryocoll}-\ref{supp:imgacqu} for details. 
$Y^1$ was then generated by applying the observation operator with known values of the model parameters to $Y^2$, and adding i.i.d. mean zero Gaussian noise. The permutation was chosen to be the identity to make it simpler to visualise a correct matching.

All test problems were evaluated through  8 independent Markov chains, on a machine with specification outlined in~\ref{supp:systemspecs}. Initial positions of chains were randomly chosen as draws from the parameter priors, and a random initial permutation vector chosen.
A minimum of $7\times10^6$ tempered samples were performed (unless stated otherwise) and a further $10^6$ samples at $T=1$, where thinned chains were used to characterise the posterior.  
The average acceptance ratio $\bar{\alpha}$ was evaluated every 2000 iterations, 
and the proposal variances adjusted accordingly to ensure efficient sampling. 

\subsection{ \textit{In silico} cell matching} \label{sec:truematches}

For the first test, we generated problems where a known random affine transformation was applied to the original $Y^2$ coordinates in order to generate $Y^1$, parameters given in Table~\ref{tab:datagenparamvalues}. Additive noise of the form $\mathcal{N}(0,0.01^2I_3)$ was then added to each point. 

\begin{figure}[htpb]
\centering
    \includegraphics[width=0.95\textwidth]{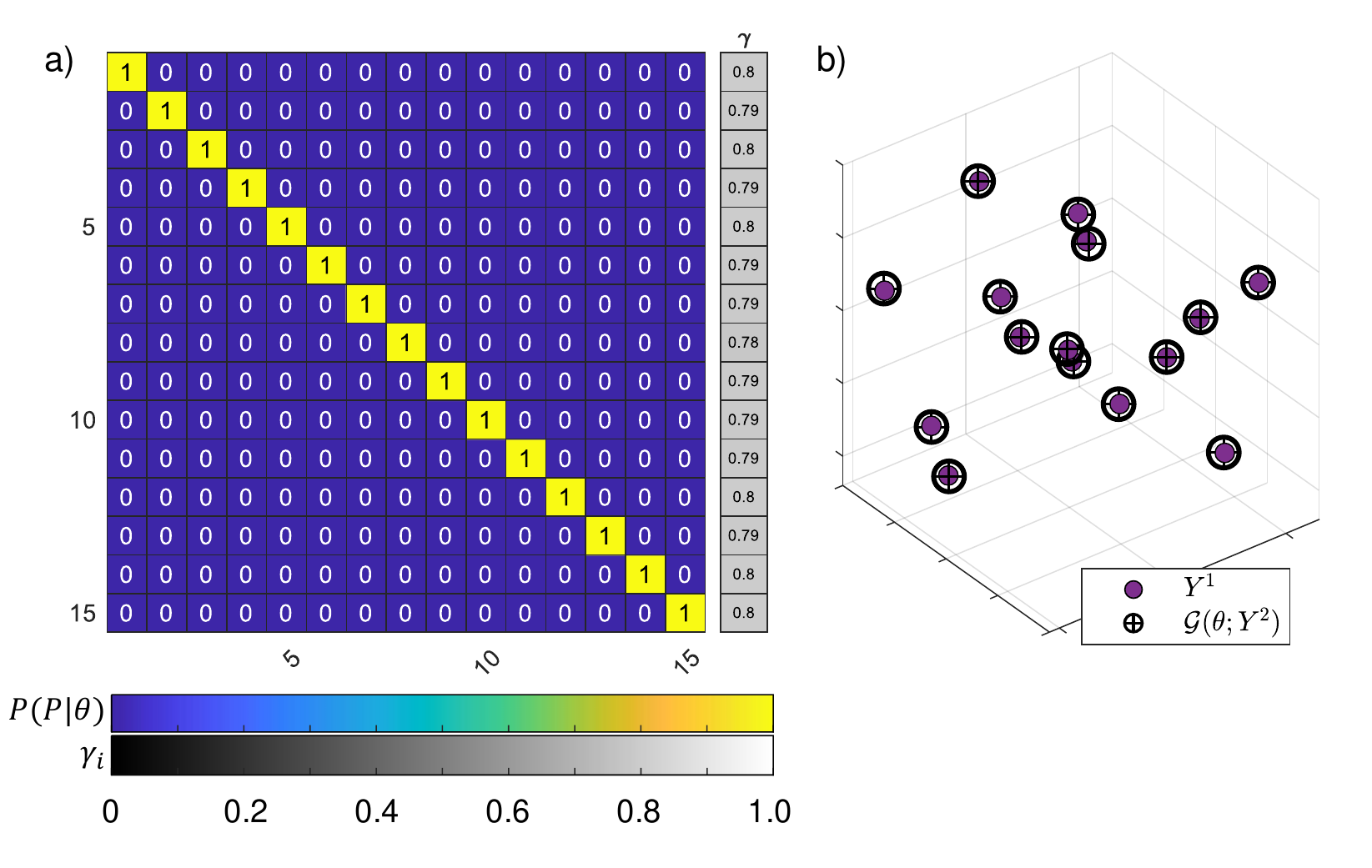}
    \caption{a) Example normalised probability heatmap of matches for the 15-cell problem with corresponding MAP estimates (conditioned on MLM) on $\gamma$. b) Corresponding spatial matching of $Y^1$ and $\mathcal{G}(\theta;Y^2)$ using estimates of transformation parameters.}
    \label{fig:15celleasytest}
\end{figure}

We performed sampling on the affine transformation parameters, the permutation vector and fidelity parameters but disregarded non-linear deformation. All chains for the 8-, 15-, and 33-cell tests converged to a posterior distribution highly concentrated on the correct matching of points as can be seen in the example permutation probability heatmap in Figure~\ref{fig:15celleasytest}a. For the 62-cell test, 7/8 chains also appeared to converge to posterior distributions highly concentrated on the correct matching of points. The outlier chain, which likely got trapped in a local minima, had an MLM with 60 incorrect matches. The incorrect matching was clearly identifiable through the median cell-to-match distances (1.077 versus an average of 0.0189 for successful chains). This indicated that harder problems with more cells may require slower tempering due to a more complex state space.

\begin{figure}[htpb]
\centering
    \includegraphics{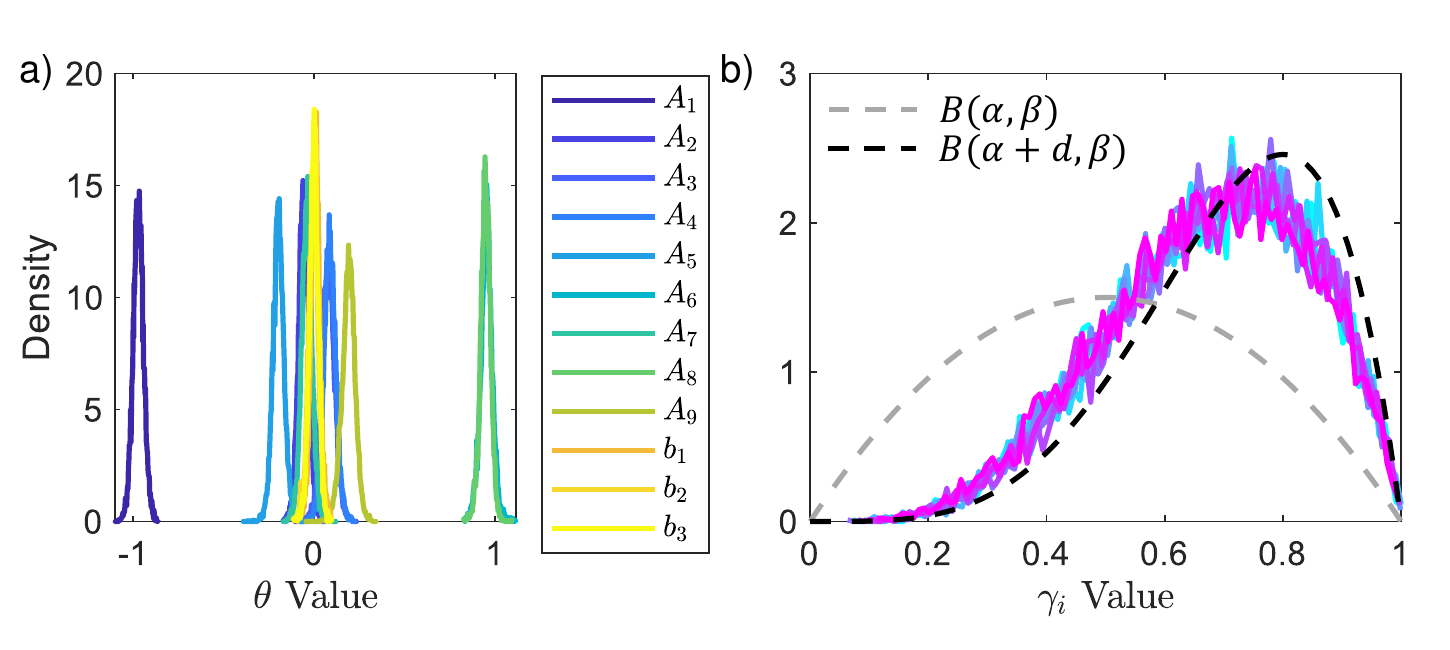}
    \caption{Marginal posteriors of a) affine transformation parameters ($A_1$-$A_9$ and $b_1$-$b_3$), b) fidelity parameters for the 8-cell \textit{in silico} example. Grey and black dashed lines are the prior and maximum possible posterior on $\gamma$ respectively.}
    \label{fig:affine/fid params}
\end{figure}

Example marginal posteriors of the affine transformation paraters $A_1$-$A_9$ and $b_1$-$b_3$ and fidelity parameters are shown in Figure~\ref{fig:affine/fid params}a,b. 
The fidelity parameter posteriors all lie close to the maximum possible fidelity posterior (corresponding to the model and data being equal), indicating excellent evidence for inclusion of all observations in this example. The noisiness of the fidelity parameter histograms is likely due to high correlations with the model parameters, causing slower convergence. 

\begin{figure}[htpb]
\centering
    \includegraphics[width=0.8\textwidth]{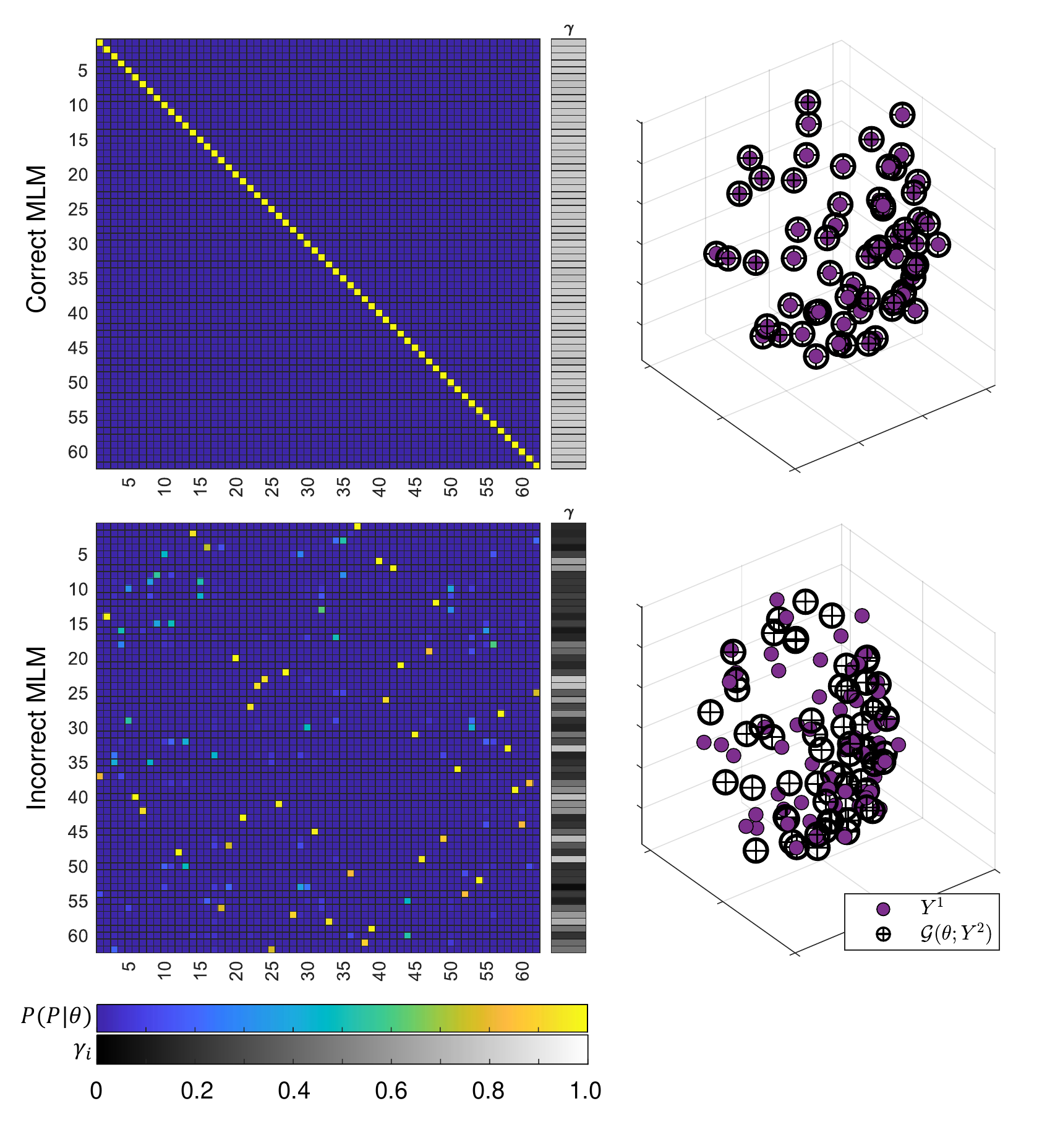}
    \caption{Comparison of two permutation heatmaps of the 62-cell problem in Section~\ref{sec:truematches}, one with the correct and one with the incorrect MLM. Clear reduction in MAP estimates, conditioned on MLM, of the fidelity parameters in the unsuccessful example, along with a reduced accuracy in the spatial matching of $Y^2$ onto $Y^1$.}
    \label{suppfig:62-celleasy}
\end{figure}

We also estimated the MAPs of the fidelity parameters conditioned on the MLM and found the fidelity parameter estimates to be close to the maximum possible posterior mode value for all cells in chains with successful identification of the MLM. The 62-cell chain where the MLM was incorrect showed significant reduction in the posterior values of the fidelity parameters for the majority of cells, further indicating the identification of a low quality, inaccurate matching, Figure~\ref{suppfig:62-celleasy}.  

In order to spatially map $Y^2$ back on to $Y^1$ and visualise the matching, we calculated the MAP estimates on the transformation parameters, conditioned on the MLM and plotted $Y^1$ and $\mathcal{G}(\theta,Y^2)$, for example Figure~\ref{fig:15celleasytest}b. There was clear correspondence of points for chains with the correct MLM as opposed to the unsuccessful chain for the 62-cell problem, Figure~\ref{suppfig:62-celleasy}.

\begin{figure}[htpb]
\centering
    \includegraphics{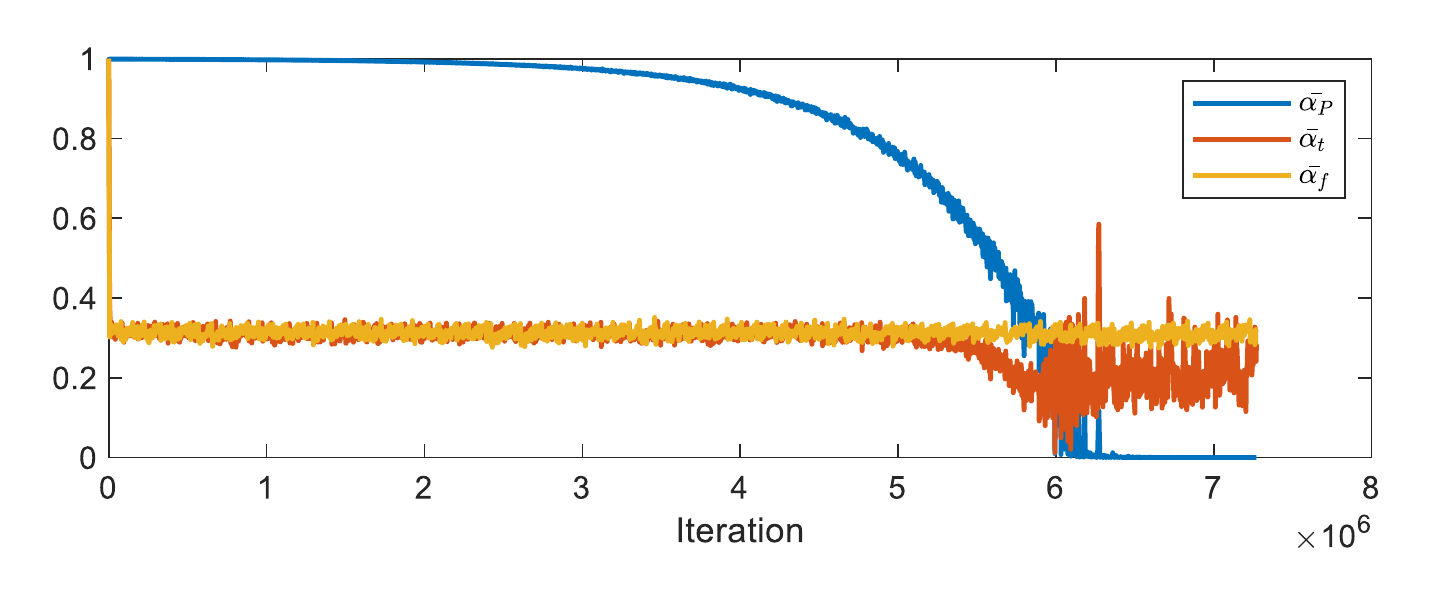}
\caption{Typical average acceptance ratio ($\bar{\alpha}$) plots for permutation, transformation and fidelity modules (subscript $P$, $t$, and $f$ respectively) during tempering.}
\label{fig:accr}
\end{figure}

Average acceptance ratios were typically stable during tempering until the average acceptance ratio of the permutation sampling, $\bar{\alpha_P}$, decreased rapidly, Figure~\ref{fig:accr}. Here the average acceptance ratio of the transformation sampling, $\bar{\alpha_t}$ fluctuated and $\beta_t$ adjusted to ensure $\bar{\alpha_t}$ was within $23.4\pm10\%$. $\bar{\alpha_P}$ was close to zero during sampling at $T=1$ for all successful chains, due to the chains likely being within the mode containing the global minimum whereby any proposed move in the permutation vector was unlikely to be accepted. 

\subsection{Data selection in presence of non-corresponding cells} \label{sec:untruematching} 
The assumption that every cell in $Y^1$ and $Y^2$ has a corresponding match does not always hold, as discussed in Section~\ref{sec:data_selection}, motivating the introduction of fidelity parameters to facilitate the selection of data within the point sets. If there is sufficient evidence that a match can not be described by the current model, the fidelity parameter posterior will have small mean, dramatically reducing the impact of that observation on the likelihood.

To investigate the effectiveness of Bayesian data selection in an \textit{in silico} setting, we simulated two test problems based on the 33- and 62-cell embryos. As before, we applied a random affine transformation, parameter values given in  Table~\ref{tab:datagenparamvalues}, and added noise of the form $\mathcal{N}(0,0.01^2I_3)$ to each point. To introduce cells without corresponding matches whilst maintaining $n_1=n_2$, we removed the first $n_r$ cells from $Y^1$ and the last $n_r$ cells from $Y^2$, resulting in $n_r$ cells in $Y^1$ and $Y^2$ without corresponding matches. We first generated two problems where $n_r=3$, and $6$ for the 33- and 62-cell data sets respectively. We chose to model these two stages as cells divide asynchronously at this stage in development, making the presence of points without associated matches more likely. For now we neglect the non-linear deformation. 

\begin{figure}[htpb]
\centering
    \includegraphics{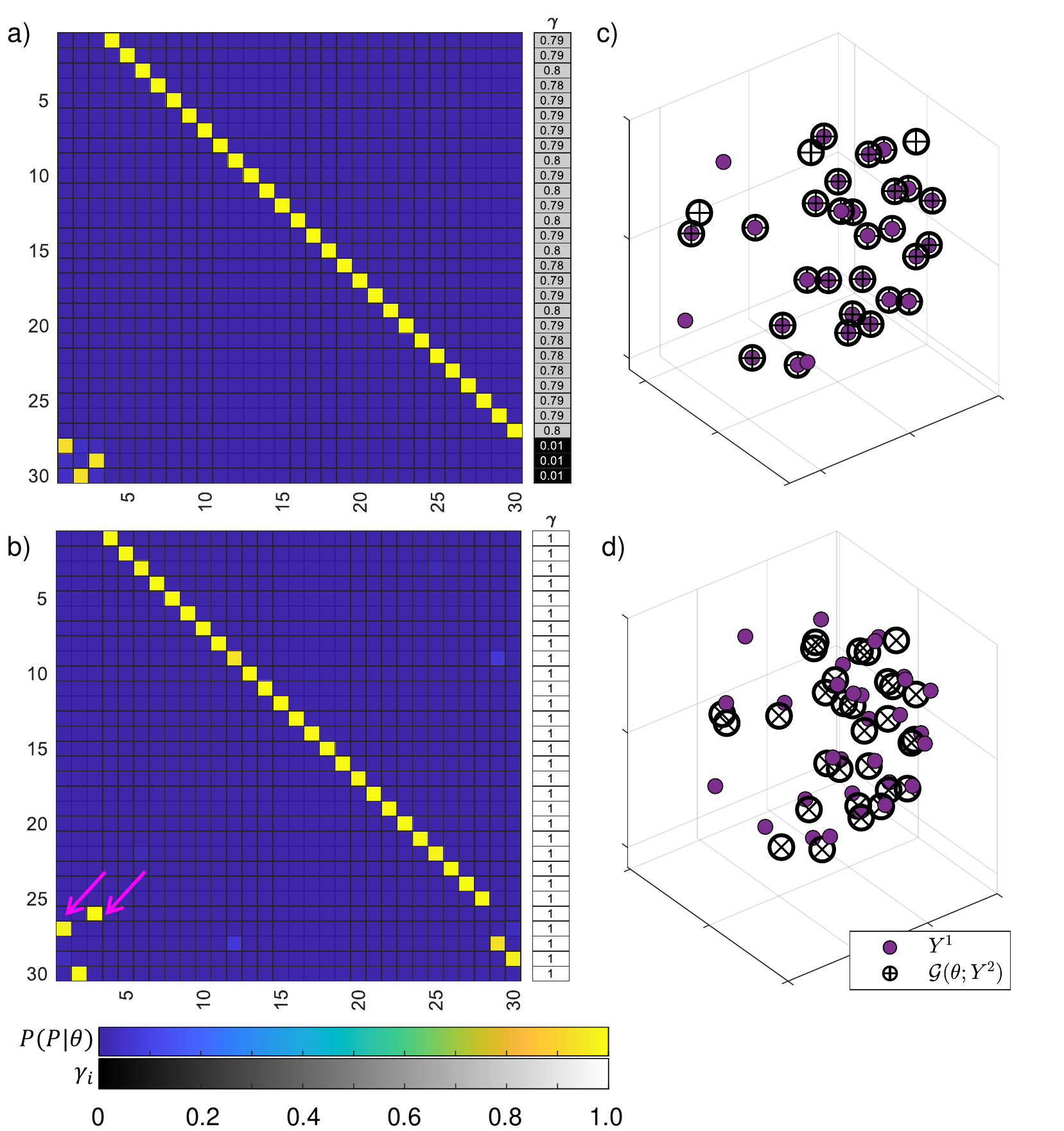}
    \caption{Comparison of matching for 33-cell test with $n_r$ cells without corresponding matches. a) Example of a permutation heatmap when data selection was included, with associated MAPs of $\gamma$, conditioned on MLM.  b) Corresponding heatmap when data selection was not included, with two incorrect matches (pink arrows). c-d) Spatial matching of points for example with and without data selection respectively.}
    \label{fig:fid10perc33-cell}
\end{figure}

All chains for the 33-cell tests converged to distributions which were highly concentrated on the correct permutation vector, with reductions in the final $n_r$ fidelity parameter posterior distributions, Figure~\ref{fig:fid10perc33-cell}a. The MLM identified was the expected permutation vector for the first $1\rightarrow (n_1-n_r)$ cells in $Y^1$, with the final $n_r$ cells matching to cells $1\rightarrow n_r$ in $Y^2$ as expected.

We then compared these results with examples where we did not include data selection. All 8 chains in the 33-cell example were concentrated about an MLM with 2 incorrect matches, Figure~\ref{fig:fid10perc33-cell}b.

We compared the median and RMSE cell-to-match distances with and without data selection, see Tables~\ref{tab:10percdist}, \ref{tab:fid_all_with} and \ref{tab:fid_all_without}. It was evident that at a small cost to the RMSE, we were able to reduce the median cell-to-match distance, thereby facilitating a better, more accurate matching for the majority of cells with definitive matches, as can be seen in Figure~\ref{fig:fid10perc33-cell}c-d. Without data selection, the matching identified is the effective result of minimising the RMSE of the cell-to-match distances for all cells, including those without corresponding matches. 

\begin{table}[htbp]
\centering
\begin{tabular}{l|cccc|}
\cline{2-5}
                                      & \multicolumn{2}{c|}{With data selection}                & \multicolumn{2}{c|}{Without data selection} \\ \cline{2-5} 
                                      & \multicolumn{1}{c|}{median} & \multicolumn{1}{c|}{RMSE} & \multicolumn{1}{c|}{median}     & RMSE      \\ \hline
\multicolumn{1}{|l|}{33-cell $n_r=3$} & \multicolumn{1}{c|}{0.0293} & \multicolumn{1}{c|}{2.03} & \multicolumn{1}{c|}{0.666}      & 1.78     \\ \hline
\multicolumn{1}{|l|}{62-cell $n_r=6$} & \multicolumn{1}{c|}{0.0226} & \multicolumn{1}{c|}{1.81} & \multicolumn{1}{c|}{0.685}      & 0.652     \\ \hline
\end{tabular}
\caption{Example of median and RMSE cell-to-match distances corresponding to the chains with the minimum values of the negative log of the posterior.}
\label{tab:10percdist}
\end{table}

\begin{figure}[htpb]
\centering
    \includegraphics[width=\textwidth]{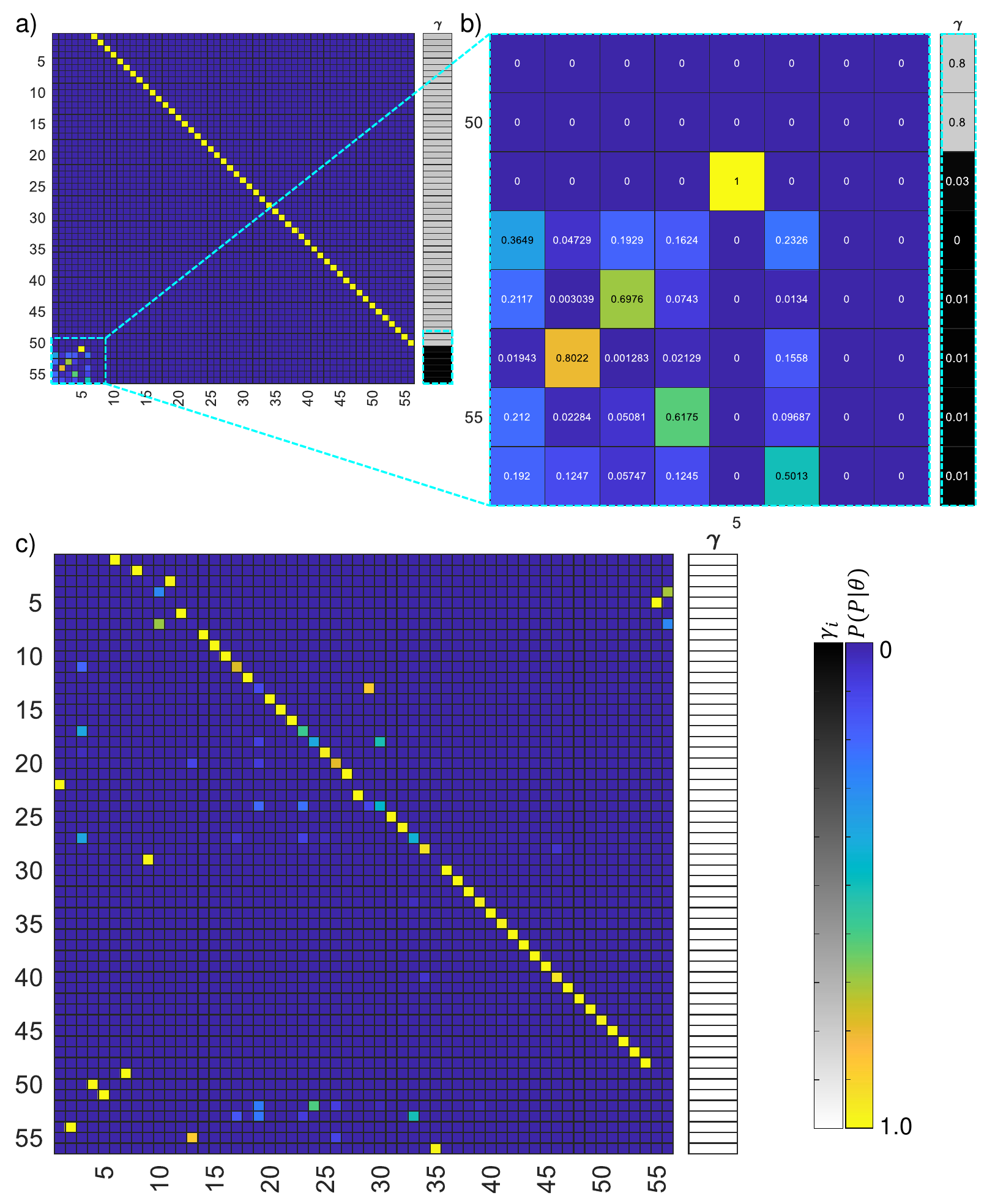}
    \caption{Example of the 62-cell test with $n_r=6$ non-corresponding cells from Section~\ref{sec:untruematching}. a) Permutation heatmap when data-selection is included. b) Inset of region where cells have no corresponding matches and reduced $\gamma$. c) Example permutation heatmap when data selection is not included, 14 incorrect matches in the MLM. Less concentrated distribution about the MLM.}
    \label{suppfig:fid10perc62-cell}
\end{figure}

Larger problems with more densely packed points could result in an increased number of incorrect matchings, as we found in the 62-cell example with $n_r=6$. There were between 8 and 50 incorrect matches in the MLM when data selection was not included, and the distribution appeared less concentrated on the correct permutation vector in all chains,  Figure~\ref{suppfig:fid10perc62-cell}c and Tables~\ref{tab:fid_all_with} and \ref{tab:fid_all_without}. This variability in the number of errors is indicative of a posterior that is much more difficult to explore, leading to local trapping. When we did include data selection, we were able to retrieve an MLM equal to the true matching in all chains. In this instance, Bayesian data selection helped us not only identify suitable data to be registered between $Y^1$ and $Y^2$, but also to smooth the posterior making it easier to explore.

In the 62-cell test problem we observed an increase in the RMSE of cell-to-match distances when data selection was included, but improvement in the median cell-to-match distance, indicative of an improved matching of the majority of cells, see tables~\ref{tab:fid_all_with} and~\ref{tab:fid_all_without}. We also noted that the cells with reduced posterior means of their fidelity parameters (the final $n_r$ cells in $Y^1$) displayed non-committal matching; matches with low fidelity could be freely exchanged,  Figure~\ref{suppfig:fid10perc62-cell}a-b. We conducted a test with larger values for $n_r$ with even more stark differences in success, see~\ref{supp:fid20perc}.

\subsection{Non-linear deformations} \label{sec:deftests}

We next sought to incorporate non-linear deformation within the data. We generated a test problem based on the 33-cell data set where we assigned non-zero momenta, drawn from the prior, to 18 points where $x<0$, and deformed explicitly through equations~\ref{eq:deftosolve_pos} and~\ref{eq:deftosolve_mom}. The points were then subject to an affine transformation, all parameters  given in Table~\ref{tab:datagenparamvalues}. Noise of the form $\mathcal{N}(0,0.01^2I_3)$ was then added. We designed four tests covering all combinations of inclusion of deformation in the observation operator and/or data selection.

When neglecting non-linear deformation and data selection, referred to as test a), we found that all chains had the same MLM with two incorrect matches, see table~\ref{tab:deftestsummary}. Although the number of errors in this particular example is low, when we trialled another test problem with the initial momenta scaled by a factor of 1.1, we found four unique MLMs with up to 33 incorrect matches. Without data selection and the inclusion of the non-linear deformation, even small increases in problem difficulty can lead to large numbers of incorrect matches.

Next we included non-linear deformation within $\mathcal{G}$, but neglected data selection, test b). The 8 chains converged to different regions, with 8 unique MLMs and between 0 and 19 incorrect matches. The posterior  here is higher dimensional and more complex, leading to poor mixing. Interestingly, when we compared the minimum negative log of the posterior density after optimisation conditioned on the MLM, we found that the chain with the deepest mode actually corresponded to a chain with 4 incorrect matches \emph{not} the chain with all correct matches, see table~\ref{tab:deftestsummary}. Mirroring this, the median and RMSE cell-to-match distance was in fact larger for the chain with 0 incorrect matches when compared to the chain with 4 incorrect matches, see table~\ref{tab:deftestsummary}, suggesting this was a fluke, and would actually lead to selection of the incorrect matching. Without the fidelity parameters to assist in interpretation of the results, the identification of good matches is ambiguous.

We then included data selection and non-linear deformation within $\mathcal{G}$, test c) and observed similar results as in test b), with an obviously multi-modal, difficult to sample from posterior distribution, resulting in the identification of 8 unique MLMs. The MLMs had between 2 and 32 incorrect matches, see table~\ref{tab:deftestsummary}, again indicating poor mixing of the chains. This test had the additional difficulty that the prior on the momenta must be carefully balanced with the prior on the fidelity parameters.

\begin{figure}[htpb]
    \centering
    \includegraphics{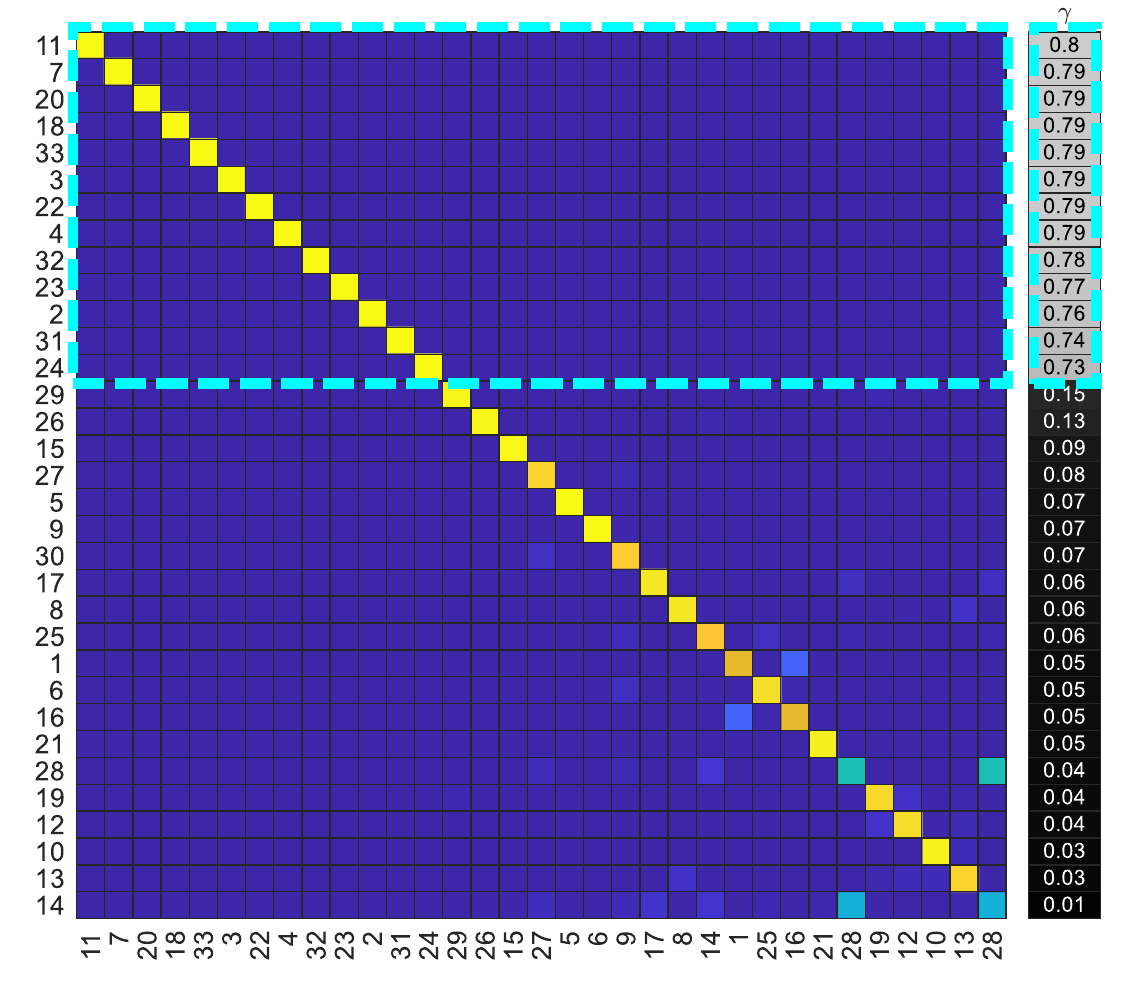}
    \caption{Permutation heatmap for the non-linear deformation test d). Cells ordered vertically and horizontally according to the ordering of the MAP estimates of the fidelity parameters conditioned on the MLM. Undeformed cells highlighted with cyan box.}
    \label{fig:deftest}
\end{figure}

We are most interested in the identification of cell matchings where we are confident in the identified matching, i.e. not necessarily identifying \emph{all} cells' matches. We therefore neglected non-linear deformation but included data selection, test d). We found that the 8 chains identified 2 unique MLMs with either 5 or 6 incorrect matches. The cells with incorrect matches were associated with reduced fidelity parameter posterior means ($<0.15$) and corresponded to cells which were explicitly deformed in the generation of the test problem. We identified consistent matching for cells with MAP estimates of fidelity parameters (conditioned on the MLM) greater than 0.5, see Figure~\ref{fig:deftest}, which corresponded with cells from the undeformed region. Two cells that were not deformed explicitly did have reduced posterior means of their fidelity parameters, but this is due to the interaction of points and their mutual repulsion via $\sigma_K$ in equations~\ref{eq:deftosolve_pos} and~\ref{eq:deftosolve_mom}.

We compared the median and RMSE cell-to-match distances for test d) with the previous tests and found that they were higher. However, when we considered only the undeformed cells, we found that the median cell-to-match distances were greatly reduced, indicating a successful matching of this subset of undeformed cells, see table~\ref{tab:deftestsummary}. We also trialled the more difficult test where the initial momentum was scaled by a factor of 1.1, and sampled only on the affine transformation, permutation vector and fidelity parameters. The 8 chains identified one unique MLM, and all cells that were subject to an initial deformation had low posterior means of their fidelity parameters indicating the successful reduction of their contribution to the likelihood. As for the previous example, we observed reduction of the median cell-to-match distance for the undeformed cells, with correct matchings, again suggesting a good matching for the subset of undeformed cells.

A final key point regarding the benefit of including data selection rather than complex non-linear deformation models is the significant improvement in run-time, due to the cost of solving the ODEs \eqref{eq:deftosolve_pos}-\eqref{eq:deftosolve_mom}. Tests b) and c) that included the deformation took approximately 28 hours to run, and suffered from slow mixing due to additional dimensionality and complexity of the posterior. On the other hand, test d) took approximately 30 minutes and converged to consistent MLMs therefore making it a far more feasible approach to match subsets of cells accurately within reasonable time frames. 

\subsection{Validation of cell matching for fixed embryos using reference markers} \label{sec:fixedtofixed}

Next we devised a simple biological test problem where we introduced reference markers within the embryo via microinjection. We collected embryos at the 8-cell stage and then microinjected a single cell with H2b-mCherry, a fluorescent protein. Embryos were then subject to 24 hours \textit{ex vivo} culture and then fixed and stained with Hoechst to facilitate nuclear segmentation. See Sections \ref{supp:embryocoll}-~\ref{supp:imgacqu}, ~\ref{supp:microinj}, ~\cite{alithesis2017} and ~\cite{Plusa2005} for full protocols. 

\begin{figure}[htpb]
    \centering
    \includegraphics{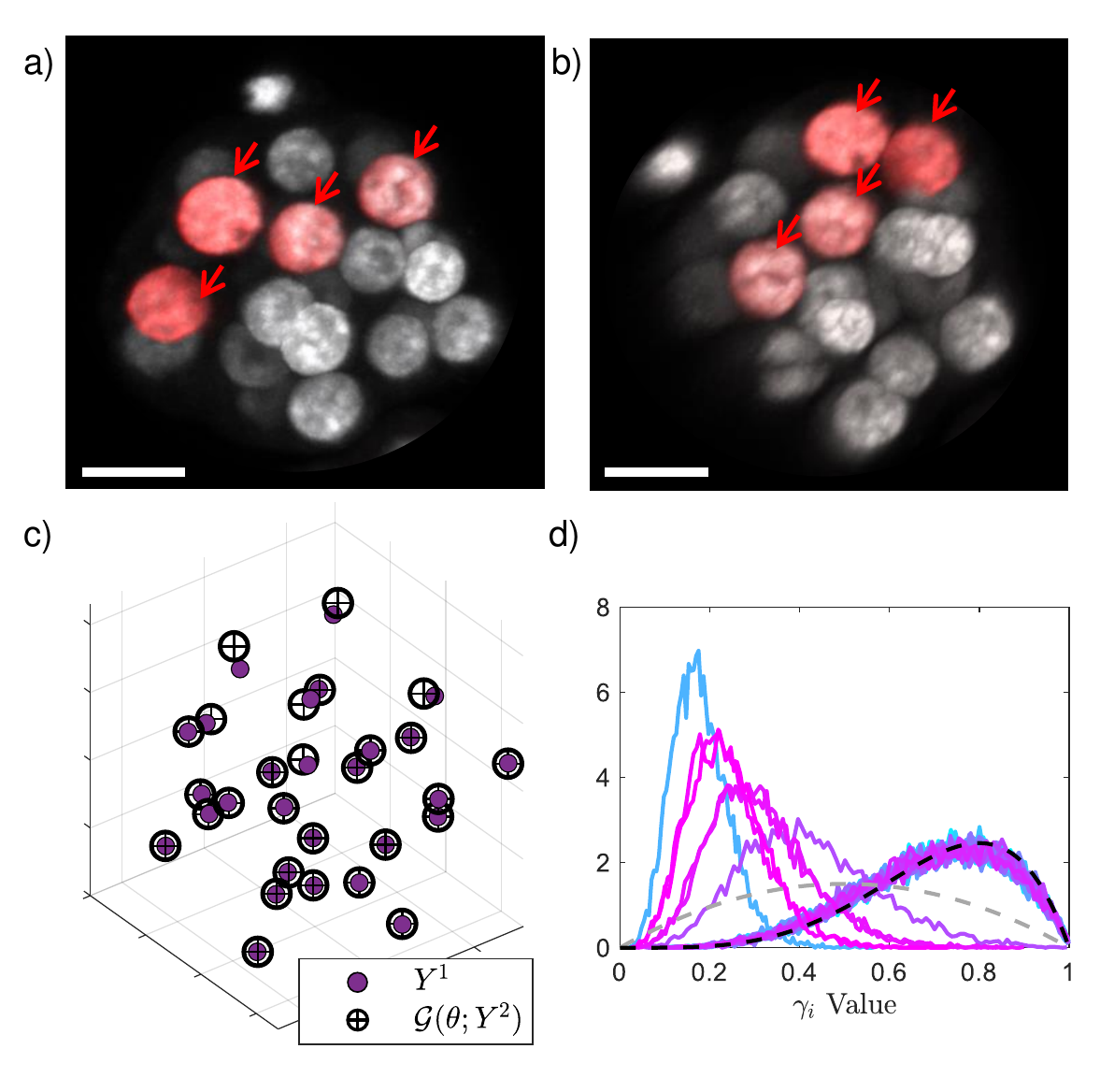}
    \caption{a) The first confocal image of the embryo prior to movement on the imaging stage- reference cells marked with arrows b) Second image of embryo after manipulation. Scale bar equal to 20$\mu m$. c) Spatial matching of points using MAP estimates of transformation parameters. d) Example of marginal posteriors on the fidelity parameters, with maximum possible posterior in black dashed line.}
    \label{fig:fixedtofixed}
\end{figure}

We selected one embryo where four mCherry positive cells were identified and used as reference markers. The embryo was imaged and then moved randomly using a pipette before a second image of the embryo was taken, Figure~\ref{fig:fixedtofixed}a-b. Cell centres were approximated through segmentation of the nuclei in both images, see Section \ref{supp:imgacqu}. 

We performed inference including data selection but neglecting non-linear deformations, and initiated 8 chains randomly using draws from the priors. A minimum of $N=10^7$ tempered iterations were conducted, and a further $N_1=10^6$ iterations at $T=1$. 

All eight chains were found to have the same MLM and had good spatial matching between the two point sets with an average median cell-to-match distance equal to 0.0403 units across the 8 chains, Figure~\ref{fig:fixedtofixed}c. We noticed that 6 cells had reduced fidelity posterior means in this example, Figure~\ref{fig:fixedtofixed}d, but not so low as to indicate a poor overall matching.

\begin{figure}[htpb]
    \centering
    \includegraphics{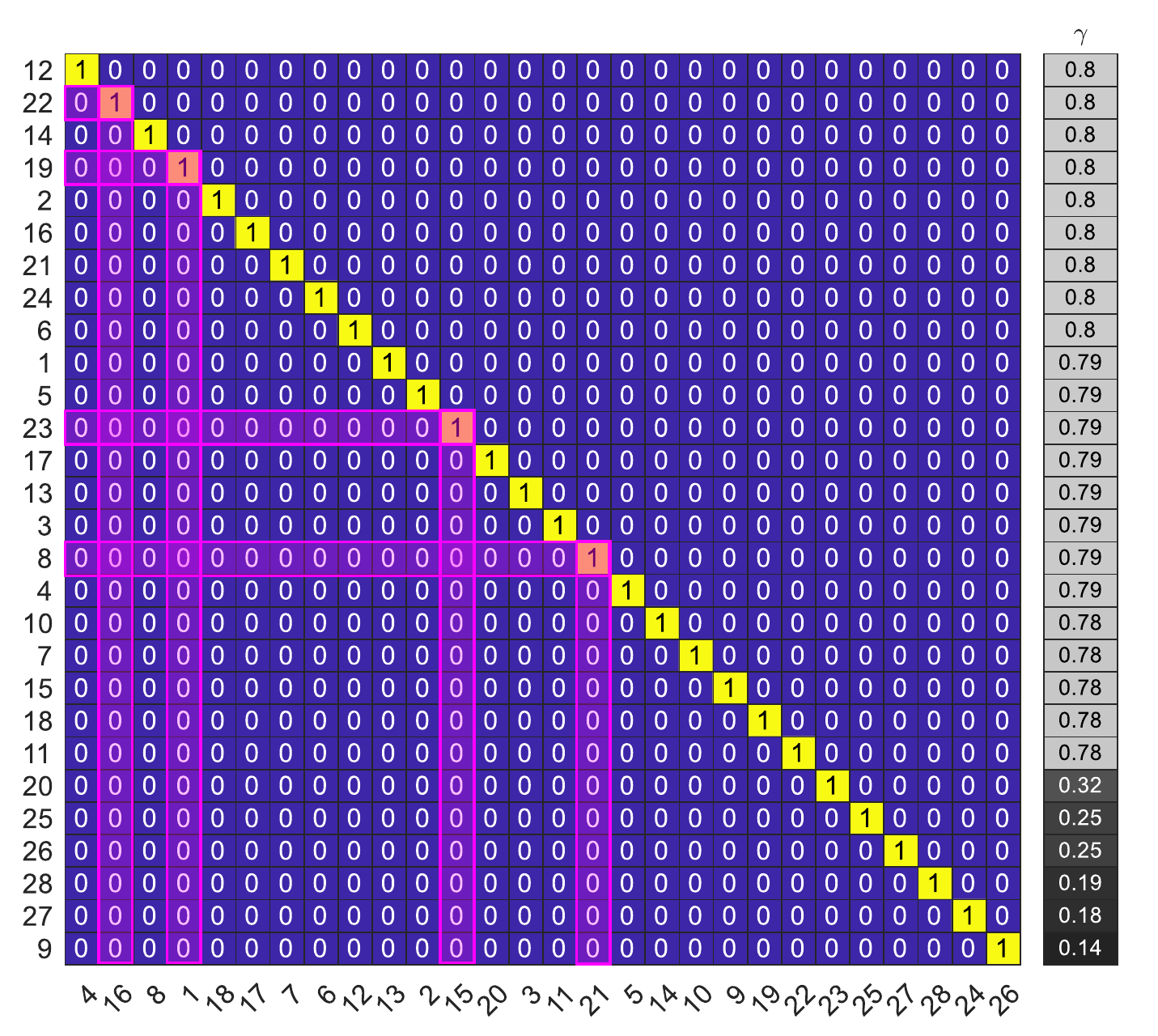}
    \caption{Example permutation heatmap from the problem in Section~\ref{sec:fixedtofixed} with four known mCherry reference markers, ordered by fidelity parameter and cell matches in $Y^2$. highlighted rows/columns indicate the successful matching of four reference maker cells.}
    \label{suppfig:fixedtofixed}
\end{figure}

Previously, in our \textit{in silico} tests we knew the ground truth permutation of the points and were able to construct heatmaps which approximated a diagonal matrix. In real data tests we no longer know this ordering \emph{a priori} for the purposes of visualisation. Therefore the ordering of cells in the permutation heatmap was arranged such that the cells in $Y^1$ were ordered according to the MAP estimates of the fidelity parameters (conditioned on the MLM), and then the order of $Y^2$ changed according to the maximum match probability for each cell in $Y^1$. The resulting heatmap was a diagonal matrix and we were able to show that the reference cells in $Y^1$ corresponded to the reference cells in $Y^2$,  Figure~\ref{suppfig:fixedtofixed}. 

\subsection{Matching of cells and embryos across imaging modalities}\label{sec:movietofixed}

Finally, we wanted to trial matching cells between the final frame of a RTI experiment and an immunostained image. H2b:GFP embryos were chosen to facilitate the segmentation of cell centres from the movie, and were subject to \textit{ex vivo} culture. Prior to removal of the embryos from the confocal microscope, they were imaged a final time using a $z$ resolution of 1$\mu$m to increase the accuracy of the extracted cell centres. Embryos were then fixed to halt development and stained using Hoechst to enable visualisation of the nuclei for segmentation. Details of experimental protocol given in Section \ref{supp:embryocoll}-~\ref{supp:rti}. 

\begin{figure}[htpb]
    \centering
    \includegraphics[]{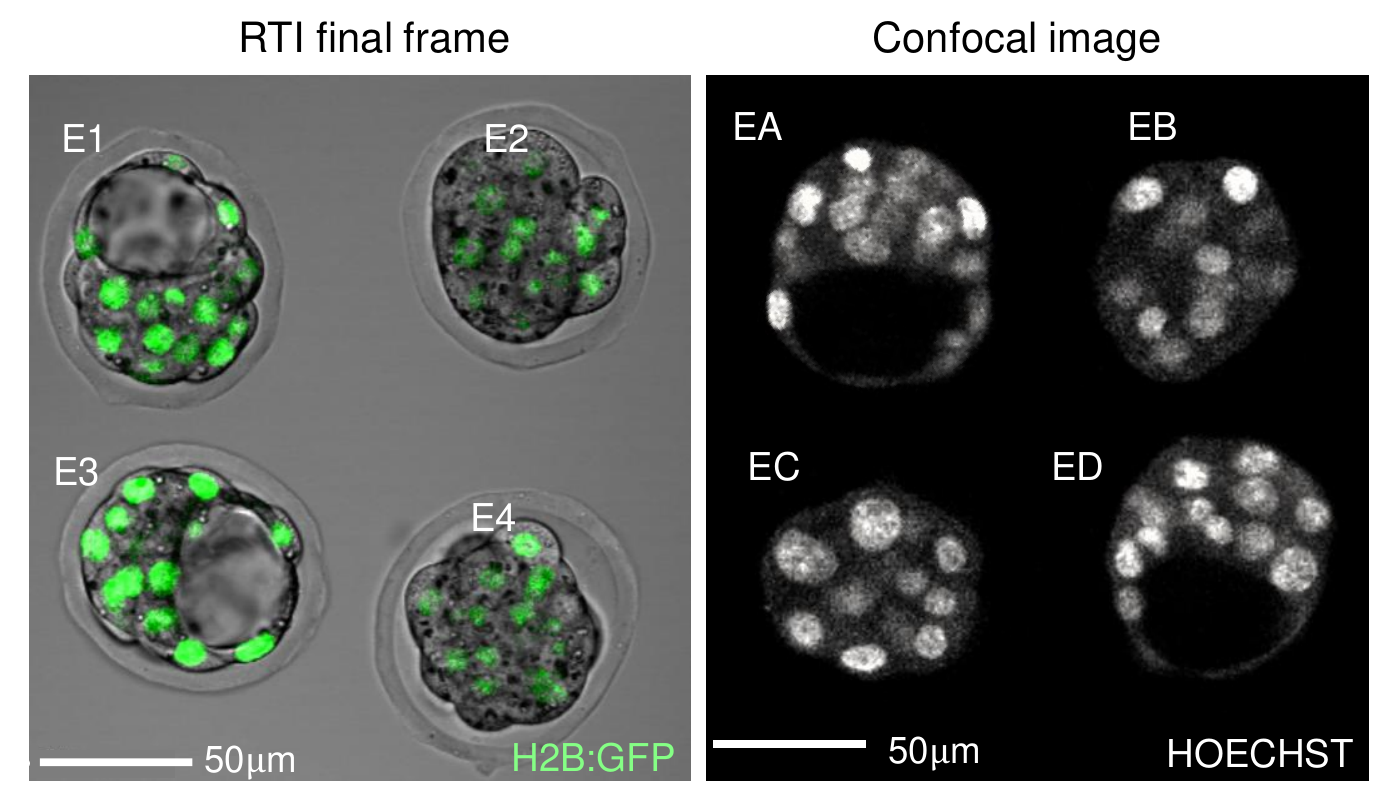}
    \caption{Representative 2D slices from RTI study and immunostaining, confocal image. Embryos 1-4 from the final frame of the RTI, with cell nuclei visualisation via the green fluorescent channel (H2b:GFP signal). Corresponding immunostained, confocal image, embryos A-D with nuclei visualisation via Hoechst staining.}
    \label{fig:movietofixedbio}
\end{figure}

We chose a group of four embryos (embryos 1-4) that were co-cultured and successfully stained (embryos A-D). Due to the co-culture of the embryos, the embryo matching was unknown \textit{a priori}, Figure~\ref{fig:movietofixedbio}. Embryos 1-4 had 39, 22, 37 and 28 cells respectively, and embryos A-D had 39, 23, 27 and 40 cells respectively. Each embryo combination was attempted (8 chains for each combination) using data selection but no non-linear deformation. We ran a minimum of $N=10^7$ tempered iterations and a further $N_1=10^6$ iterations at $T=1$.

\begin{figure}[htpb]
    \centering
    \includegraphics{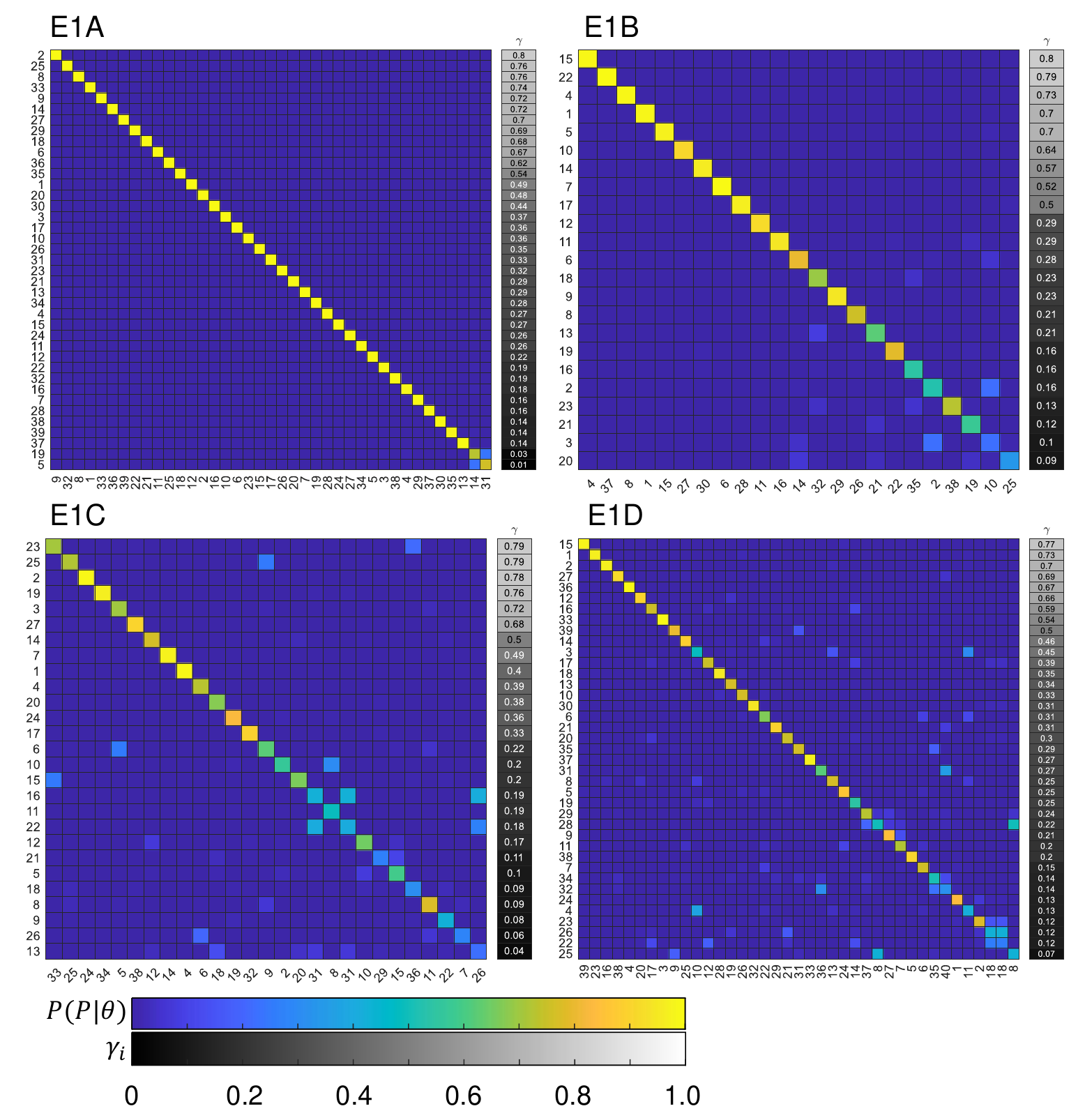}
    \caption{Example permutation heatmaps for embryo 1 with embryos A, B, C and D from Section~\ref{sec:movietofixed}. Heatmaps ordered by MAP estimate conditioned on MLM of fidelity parameters and then the corresponding maximum match in $Y^2$.}
    \label{suppfig:em1example}
\end{figure}

We noticed that for each of the embryo combinations, one of the embryo pairings would typically have a more concentrated posterior distribution about some permutation vector, whereas the remaining three embryo combinations would show more disperse distributions, with a larger variability in the permutation vector, see  Figure~\ref{suppfig:em1example}. We also noticed that the 8 chains would converge to the same MLM when the optimum embryo matching was identified for embryos 1, 2 and 4 but would otherwise identify at least 3 unique MLMs, see table~\ref{tab:uniqueMLMs}.

\begin{table}[htbp]
\centering
\begin{tabular}{c|c|c|c|c|}
\cline{2-5}
                                   & \textbf{emA} & \textbf{emB} & \textbf{emC} & \textbf{emD} \\ \hline
\multicolumn{1}{|c|}{\textbf{em1}} & 1            & 8            & 6            & 7            \\ \hline
\multicolumn{1}{|c|}{\textbf{em2}} & 7            & 1            & 3            & 6            \\ \hline
\multicolumn{1}{|c|}{\textbf{em3}} & 3            & 6            & 8            & 5            \\ \hline
\multicolumn{1}{|c|}{\textbf{em4}} & 6            & 7            & 1            & 7            \\ \hline
\end{tabular}
\caption{Number of unique MLMs identified for each embryo combination, out of 8 chains. }
\label{tab:uniqueMLMs}
\end{table}

\begin{figure}[htpb]
    \centering
    \includegraphics{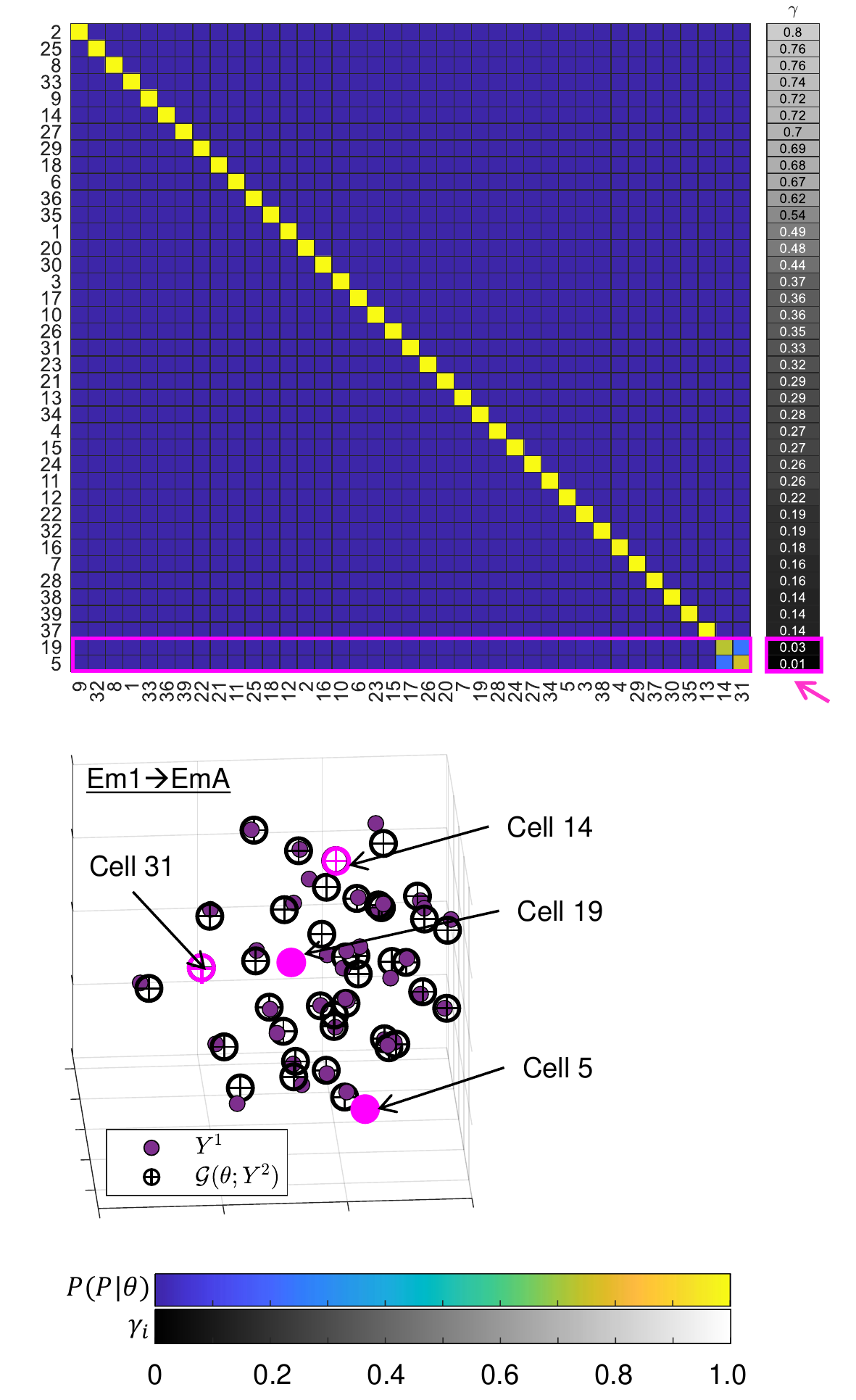}
    \caption{a) Example permutation heatmap for the matching of embryo 1 with embryo A in Section~\ref{sec:movietofixed}. Two cells with significantly reduced fidelity parameter MAP estimates conditioned on the MLM highlighted in pink box. b) Spatial mapping of $Y^1$ onto $Y^2$ with low fidelity cells marked in pink. Cells with low fidelity parameters found in regions deep in the embryo and in the extremes of the z axis where segmentation errors are more likely to occur.}
    \label{suppfig:movietofixed}
\end{figure}

As before, we recorded the cell-to-match distances during sampling at $T=1$ and then compared the median and RMSE distances from the chain that converged to the minimum negative log of the posterior for each embryo pairing, see table~\ref{tab:movietofixed}. By considering the median cell-to-match distance, we were able to clearly identify three embryo matchings; embryo 1, 2 and 4 with A, B and C respectively. Interestingly, these embryo pairings did not always correspond to the lowest RMSE distances. For instance, the pairing of embryo 1 with embryo A had the third largest RMSE distance, despite having the overall minimum median cell-to-match distance. On closer inspection, this was due to the reduction of several fidelity parameters. We noticed that cells 5 and 19 in $Y^1$ had significantly reduced posterior means of the corresponding fidelity parameters and were matched to cells 14 and 31 in $Y^2$ causing the increase in the RMSE, see Figure~\ref{suppfig:movietofixed}. These cells were either deep within the embryo or in the extremes of the z axis, both regions where segmentation can be prone to errors. The identification and description of these points via the fidelity parameters, could help us go back and re-segment these regions more accurately, something that without data selection we would not be able to do.

\begin{table}[htbp]
\centering
\begin{tabular}{c|c|c|c|c|lc|c|c|c|c|}
\cline{2-5} \cline{8-11}
\multicolumn{1}{l|}{Median}            & \textbf{emA} & \textbf{emB} & \textbf{emC} & \textbf{emD} &                       & \multicolumn{1}{l|}{RMSE} & \textbf{emA} & \textbf{emB} & \textbf{emC} & \textbf{emD} \\ \cline{1-5} \cline{7-11} 
\multicolumn{1}{|c|}{\textbf{em1}} & \textbf{0.359}        & 0.645        & 0.809        & 1.316        & \multicolumn{1}{l|}{} & \textbf{em1}            & \textbf{1.338}        & 0.771        & 1.012        & 1.366        \\ \cline{1-5} \cline{7-11} 
\multicolumn{1}{|c|}{\textbf{em2}} & 0.696        & \textbf{0.233}        & 0.632        & 0.730        & \multicolumn{1}{l|}{} & \textbf{em2}            & 0.771        & \textbf{0.180}        & 0.617        & 0.812        \\ \cline{1-5} \cline{7-11} 
\multicolumn{1}{|c|}{\textbf{em3}} & 1.515        & 0.846        & 0.879        & 0.775        & \multicolumn{1}{l|}{} & \textbf{em3}            & 1.224        & 0.909        & 1.039        & 0.923        \\ \cline{1-5} \cline{7-11} 
\multicolumn{1}{|c|}{\textbf{em4}} & 1.032        & 0.765        & \textbf{0.333}        & 0.755        & \multicolumn{1}{l|}{} & \textbf{em4}            & 0.793        & 0.743        & \textbf{0.310}        & 0.747        \\ \cline{1-5} \cline{7-11} 
\end{tabular}
\caption{Median and RMSE cell-to-match distances for each embryo combination, given in arbitrary units corresponding to the chain that converged to the minimum negative log of the posterior density. Bold entries correspond to high confidence embryo matchings.}
\label{tab:movietofixed}
\end{table}

\begin{figure}[htpb]
    \centering
    \includegraphics{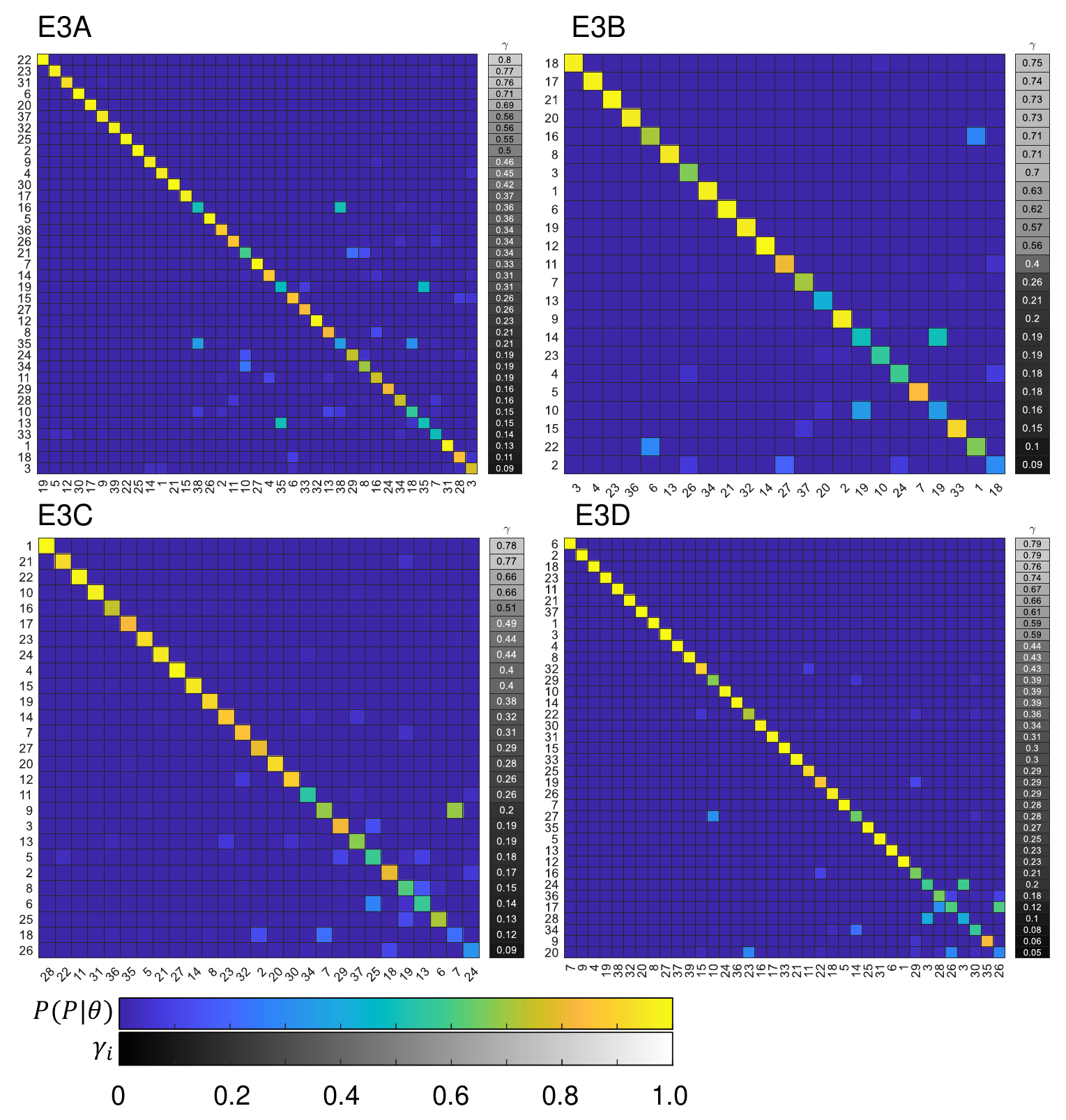}
    \caption{Example permutation heatmaps for embryo 3 with embryos A, B, C and D from Section~\ref{sec:movietofixed}. Heatmaps ordered by MAP estimate conditioned on MLM of fidelity parameters and then the corresponding maximum match in $Y^2$.}
    \label{suppfig:movietofixed3}
\end{figure}

By deduction we can say that embryo 3 should match with embryo D, however this was not as clear through assessment of the median cell-to-match distance or permutation heatmaps as for the other embryos. The 8 chains converged to three unique MLMs, again increasing our uncertainty in the identified matchings. The permutation heatmaps were typically more diffuse for all embryo combinations, suggesting that we were unable to identify a single global minimum indicative of the true cell matching, see Figure~\ref{suppfig:movietofixed3}. The inaccuracy of the matching between embryo 3 and D could potentially be due to some increased deformation during fixation or inaccuracies during segmentation of the cell centres. This highlights the requirement for good quality data when matching the points. 

For comparison, we performed the cell matching for each of the well-identified embryo matches (embryo pairs 1A, 2B and 4C) without data selection. All 8 chains for embryo pairings 2A and 4C identified the same MLM as with data selection, as we would hope high quality data. However, when we tried to match embryo 1 with embryo A without data selection, we identified MLMs with large numbers of differences when compared to the MLM identified previously with data selection. One chain had 7 differences, 6 of the chains had 11 differences and the final chain had 39 differences, indicating a completely different identified MLM. This highlights our need to include the data selection framework, to ensure the accurate matching where there are cells without a corresponding match. Furthermore, the inclusion of data selection facilitates further inference and interpretation of the confidence of the matches presented within the MLM, and enables better mixing of the Markov chains due to its smoothing properties.

\section{Discussion}  \label{sec:discus}

In this work we presented a solution to an unlabelled landmark registration problem,  by introducing a novel Bayesian data selection approach to account for non-corresponding cells. We included non-linear deformation, 3D affine transformation and description of the matching of cells via a permutation matrix within the registration model. By using MCMC and tempering of the likelihood, we were able to explore the complex, multimodal posterior and identify most likely matchings of two point-sets. To demonstrate the efficacy of the approach, we constructed a series of \textit{in silico} problems, and used real data from biological imaging experiments. We were able to determine the matching of cells between the final frame of a RTI experiment and corresponding immunostained images, even when the embryo correspondence was originally unknown due to co-culture of the embryos.

Our development of an approach to match single cells between imaging modalities enables the combination of historical cell data extracted from RTI studies, with protein expression at the single cell level. Previously this has been approached manually, resulting in potentially subjective conclusions relating cell behaviour and protein expression. By enabling this joint assessment of spatio-temporal information at the single cell level using our approach, we can begin to investigate the importance of cell history during cell lineage specification within the mammalian preimplantation period. 

Existing landmark registration approaches are predominantly framed as optimisation problems, and therefore provide no measure of uncertainty in the identified matching of points~\cite{kent2004matching}. Some of these approaches also rely on some partial labelling of matches and additional information relating the points such as the properties of the landmarks~\cite{kent2004matching, dryden2007statistical, green2006bayesian}. In contrast our approach is based solely on the geometrical coordinates of the landmarks. 

The development of the data selection aspect of this approach was crucial to the accurate registration in the real-world problem due to the presence of cells without corresponding matches in either image. We demonstrated that without the incorporation of the data selection framework, the accuracy of identified cell matchings was reduced, especially in larger embryos where the number of cells without corresponding matches was potentially increased. We also demonstrated that the inclusion of data selection facilitated better mixing of the MCMC chains by reducing the roughness of the state space, thus improving chain convergence and improving the robustness of the approach. 
More sophisticated MCMC methods that are known to be more efficient in multimodal targets, such as parallel tempering, could be used to further improve mixing and reduce computational complexity. Choosing conjugate priors for the fidelity terms could also reduce the dimensionality of the problem, and further improve mixing. 

The idea of \emph{Bayesian data selection}, in which parameters which govern the effect of an observation on the posterior are inferred alongside the model parameters, is extremely general, with great potential to be applicable to a very broad class of inferential problems in statistics and data science. Data cleaning is a subjective and laborious task which is often undertaken by hand, the results of which can have a profound impact on the outputs of the inference, and this approach automates that process in a way which is consistent and free from user-bias. In future work we plan to explore these ideas in more depth, and apply them to a range of disparate application areas.

\section*{Acknowledgments} Thanks to Colin Cotter for useful conversations about the geodesic flows.

\section*{Funding statement}
JEF was supported by the Wellcome Trust 4 year PhD studentship (Quantitative and Biophysical Biology, 108867/Z/15/Z). AHA was supported by a scholarship from Iraqi Cultural Attaché in London (GB) (AL-ANBAKI Ref. S-1007). Collection of biological data was conducted in BP’s lab, supported by the Wellcome Trust grant Seed Awards in Science (212372/Z/18/Z). SLC was supported by the Alan Turing Institute.

\bibliographystyle{jess} 
\bibliography{refs}

\newpage

\title{Unlabelled landmark matching via Bayesian data selection, and application to cell matching across imaging modalities: Supplementary material}
\emptythanks
\maketitle

\setcounter{page}{1} 

\beginsupplement

\subsection{Embryo collection} \label{supp:embryocoll}
All embryo handling was performed as in~\cite{Forsyth2021}. Mice were collected under a 12-hour light cycle at the Biological Service Facility (BSF) at the University of Manchester. Strains used were CAG::H2B-EGFP transgenic mice to enable visualisation of the chromatin for cell coordinate extraction~\cite{Hadjantonakis2004}, or outbred CD-1 (Jackson Laboratories). Embryos were collected using the time of presence of a vaginal plug as a reference time of E0.5 (embryonic day 0.5 - 12 hours post fertilisation). Morula (pre-cavitation embryos, up to E2.5) were collected through flushing of the oviducts and post-cavitation embryos (post E3.5) collected through flushing of the uterus horns with warmed M2 medium~\cite{grabarek2012live}. Embryos that were destined for immediate fixation were subject to zona removal using warmed acid Tyrode's solution (Sigma Aldrich). After zona removal embryos were washed and left to recover in M2 on the heating stage for a minimum of 20 minutes. 

Mouse handling and husbandry followed the regulations established in the UK Home Office’s Animals (Scientific Procedures) Act 1986. The animals were bred on project license P08B76E2B, protocol 4 and the license 70/08858, protocol 4. All animals were humanely euthanised according to Schedule 1 of the UK Animals (Scientific Procedures) Act 1986. Ethical approval for the euthanasia of animals for use in this study was granted to the project submitted by Berenika Plusa by the University of Manchester Animal Welfare and Ethical Review Body on the 10/03/2017.

\subsection{Embryo fixation and immunostaining} \label{supp:fixstain}

Embryos destined for immediate staining (i.e. not subject to post-collection \textit{ex vivo} culture), were fixed in 4$\%$ para-formaldehyde (Sigma Aldrich) in PBS with 0.1$\%$ Tween-20 (Sigma) and 0.01$\%$ Triton X-100 (Fluka) for 20 minutes. Embryos were then washed and stored in PBS at 4$^{\circ}$C. 

Embryos were then subject to immunostaining using the protocol outlined in~\cite{plusa2008}. Embryos were first permeabilised in 0.65$\%$ Triton X-100 in PBS for 20 minutes and then blocked in 10$\%$ donkey serum (Sigma Aldirch) in PBS for 40 minutes at 4$^\circ$C. Only nuclear staining was required for cell segmentation for the purpose of the tests included in this work. However, the primary antibodies used for Figure~\ref{fig:methodology}, were anti-E-cadherin (Santa Cruz Biotechnology) at 1:400, anti-GATA6 (Cell Signalling Technologies) at 1:400 and anti-SOX2 (Santa Cruz Biotechnology) 1:100 overnight at 4$^\circ$C. Embryos were then washed in PBS and subject to secondary Alexa Fluor (Invitrogen) conjugated antibodies at a dilution of 1:500 in blocking buffer for 1 hour at 4°C. To visualise the nuclei and aid in nuclear segmentation, all embryos were subject to Hoechst 33342 (Sigma Aldrich) at a concentration of 1:1000 in PBX (PBS +0.1$\%$ Triton X-100) and incubated for at least 30 minutes at 4$^\circ$C.

\subsection{Image acquisition and analysis} \label{supp:imgacqu}
Embryos subject to staining or microinjection were imaged on glass-bottom dishes under mineral oil using the Nikon A1 inverted confocal microscope. Sections were imaged every 1$\mu$m in the z direction. Lasers used to excited fluorophores were Diode 405 nm, Argon 488 nm, HeNe 546 nm, and HeNe 647 nm. 
Nikon format images from the confocal microscope were then converted for analysis using the IMARIS (BitPlane) file converter. Cell nuclei centres were approximated using the spot detector within IMARIS using a minimum spot detection diameter of $6\mu m$ using the H2b:GFP or Hoechst signal. Automatically identified nuclei were then checked manually for inaccuracies such as mis-identified nuclei, or nuclei not segmented in attenuated regions of the images. Point information, including $x, y, z$ coordinates and associated signal intensities were then exported as excel files where the coordinates were copied into .txt files for importing into the matching pipeline. 

\subsection{Embryo culture and real time imaging} \label{supp:rti}
Embryos subject to RTI experiments were collected at E2.5 and cultured in homemade KSOM under mineral oil for 24 hours (until E3.5) at 37.5$^\circ$C and 5$\%$ CO$_2$~\cite{grabarek2012live}. Three-dimensional brightfield and 488nm fluorescent images of the developing embryos were acquired every 20 minutes. Images had a z resolution of $\Delta z=4\mu m$, and $\Delta x = \Delta y = 0.734\mu m$. The 488nm laser intensity was adjusted to incur minimal phototoxicity in developing embryos whilst enabling visualisation of nuclear chromatin for cell identification and tracking. 

\subsection{Description of the non-linear geodesic deformation}\label{supp:deformation}

The displacement to the cells is applied through a flow field $u_t$, which can be evaluated at the current position of the landmarks $\mathbf{q}$. The flow field advects the landmarks over the time interval $t \in [0,1]$. The flow field is chosen to be a geodesic which minimises the energy of the deformation given by
\[
\frac{1}{2}\int_0^1 \|u_t\|^2_V\, \d t,
\] which is uniquely determined by the initial momenta $p^j_t$ at $t=0$ at each landmark with coordinate $q^j$. Here $V$ is a reproducing kernel Hilbert space~\cite{younes2019} with norm $\|\cdot\|_V$, and with kernel $K_V$, which we assume to be Gaussian:
\[
K_V(\mathbf{x},\mathbf{y}) = \exp\left (-\frac{\|\mathbf{x} - \mathbf{y}\|^2_2}{2\sigma_K^2}  \right ).
\]
The geodesic deformation is then given by the solution of the following differential equations 
\begin{subequations}
  \begin{align} 
     \frac{d p_t^j}{d t} &= - \nabla(u_t(q^j_t))^\top \cdot p_t^j, \label{eq:ode_original_p}\\
     \frac{d q_t^j}{dt} &= u_t(q^j_t), \label{eq:ode_original_q}
 \end{align}
 \end{subequations}
 over the time interval $[0,1]$, where $q^j$ and $p^j$ are the initial position and momentum of the $j^{th}$ cell respectively in three dimensions. We define $u_t$ at $q^j_t$ as
 \begin{equation}
    u_t(q_t^j)= \sum_{i=1}^M K_V(q^i_t,q^j_t) p_t^i,= \sum_{i=1}^M \exp\left( -\frac{\| q_t^i - q_t^j \|^2_2}{2 \sigma_K^2} \right) p_t^i,
 \end{equation}
where $\sigma_K$ describes the variance of the kernel. As the data is pre-processed to ensure a minimum cell-to-cell distance of one, we found that $\sigma_K=1$ was a sensible value to use. Equations~\ref{eq:ode_original_p} and~\ref{eq:ode_original_q} can be re-written as 
\begin{subequations}
  \begin{align} 
         \frac{d p_t^j}{d t}  &= \left(-\sum_{i=1}^M \frac{(q_t^i-q_t^j)}{\sigma_K^2} \exp\left( -\frac{\| q_t^i - q_t^j \|_2^2}{2 \sigma_K^2} \right) p_t^i \right)^\top \cdot p_t^j\label{suppeq:deftosolve_pos},\\
         \frac{d q_t^j}{dt} &= \sum_{i=1}^M \exp\left( -\frac{\| q_t^i - q_t^j \|_2^2}{2 \sigma_K^2} \right) p_t^i. \label{suppeq:deftosolve_mom}
\end{align}
\end{subequations} 

The deformation is applied to $\mathbf{q}=Y^2$, the original positions of the cell points prior to deformation, through $\mathbf{p}$ and equations~\ref{suppeq:deftosolve_pos} and~\ref{suppeq:deftosolve_mom} solved over $t=[0,1]$ to give $\phi(\theta;Y^2)$, the deformed $Y^2$ coordinates at time $t=1$.

\subsection{Affine transformation in three dimensions} \label{supp:affine}
To apply an affine transformation to three dimensional points we first describe the matrix $A(\theta)$ as a combination of two rotation matrices $R_1(\theta)$, $R_2(\theta)$, and a scaling matrix $S(\theta)$
\begin{equation}
R_1(\theta) =  \begin{bmatrix}
	 \cos(\alpha_1) & \sin(\alpha_1) & 0\\ 
	-\sin(\alpha_1) & \cos(\alpha_1) & 0\\ 
	0 &0 &1
		\end{bmatrix}
		\begin{bmatrix}
	 \cos(\alpha_2) &0& -\sin(\alpha_2) \\ 
	0&1&0\\
	\sin(\alpha_2) &0& \cos(\alpha_2)\\ 
		\end{bmatrix}		
		\begin{bmatrix}
		1&0&0\\
	 0& \cos(\alpha_3) & \sin(\alpha_3) \\ 
	0&-\sin(\alpha_3) & \cos(\alpha_3) \\ 
		\end{bmatrix},
\end{equation}
\begin{equation}
S=\begin{bmatrix}
		s_1+1 &0 &0\\
		0 & s_2+1 & 0\\
		0& 0& s_3+1
		\end{bmatrix},
\end{equation}
\begin{equation}
R_2(\theta) =  \begin{bmatrix}
	 \cos(\beta_1) & \sin(\beta_1) & 0\\ 
	-\sin(\beta_1) & \cos(\beta_1) & 0\\ 
	0 &0 &1
		\end{bmatrix}
		\begin{bmatrix}
	 \cos(\beta_2) &0& -\sin(\beta_2) \\ 
	0&1&0\\
	\sin(\beta_2) &0& \cos(\beta_2)\\ 
		\end{bmatrix}		
		\begin{bmatrix}
		1&0&0\\
	 0& \cos(\beta_3) & \sin(\beta_3) \\ 
	0&-\sin(\beta_3) & \cos(\beta_3) \\ 
		\end{bmatrix},
\end{equation}
where $A(\theta)=R_1(\theta) S(\theta) R_2(\theta)$, $\alpha_i$ and $\beta_j$ are angles of rotation, and $s_k$ are the scaling coefficients. The translation of points is applied through 
\begin{equation}
{\mathbf{b}}(\theta)=\begin{bmatrix}
				b_1\\
				b_2\\
				b_3
				\end{bmatrix} {\mathbf{1}}_{n_2}^\top,
\end{equation}
where $\mathbf{1}_n \in \mathbb{R}^n_2$ is a column vector of ones. The affine transformation is applied as
\begin{equation}
   \mathcal{F} (\theta,Y^2)=A(\theta)\phi(\theta;Y^2) + b(\theta)\mathbf{1}_{n_2}^\top,
\end{equation}
where $\phi(\theta;Y^2)=Y^2$ when no non-linear behaviour is included within the model. 

\subsection{Calculating the $\Gamma$-dependent normalisation to the posterior density} \label{supp:datafidconstant}
The normalisation factor of the posterior distribution is no longer constant when we include data selection, instead it is dependent on $\Gamma$ as
\begin{equation}
    Z (\Gamma)= \int \left| \Psi + (X\Gamma)(X\Gamma)^\top \right| ^{-\frac{\nu + n_1}{2}} dX,
\end{equation}
We can directly calculate the $\Gamma$-dependent normalisation by considering the substitution $Y=X\Gamma$, equivalent to $y_{ij}=\gamma_i x_{ij}$. Given that $dX = \frac{dY}{|det(D_J)|}$, where $D_J$ is the Jacobian of the transformation from $X$ to $Y$, given by
\begin{equation}
    D_J = \begin{pmatrix}
\gamma_1 I_3 &  & \\
 & \ddots &  \\
 && \gamma_{n_1} I_3
\end{pmatrix}
\end{equation}
and the absolute value of the determinant given by 
\begin{equation}
    |\det(D_J)|= \prod_{i=1}^{n_1} \gamma_i^d.
\end{equation}
From this we can re-write $dX$ and write $Z(\Gamma)$ as a combination of a $\Gamma$ dependent function multiplied by some constant
\begin{subequations}
    \begin{align}
        Z (\Gamma) &= \left(\prod_{i=1}^{n_1}  \gamma_i^{-d} \right) \int \left| \Psi + YY^\top \right| ^{-\frac{\nu + n_1}{2}} dY, \\
        &= \left( \prod_{i=1}^{n_1} \gamma_i^{-d} \right) \frac{\pi^{d n_1/2} \Gamma_d(\frac{\nu}{2})}{|\Psi|^{\nu/2} \Gamma_d(\frac{\nu +n_1}{2})}.
    \end{align}
\end{subequations}
By dropping the constant terms in $Z (\Gamma)$ and retaining only the factor dependent on $\Gamma$, we re-write the posterior as
\begin{equation} 
\pi(\theta, \Gamma | Y^1, Y^2) \propto \pi_0(\theta) \pi_0(\mathbf{\gamma}) \left( \prod_{i=1}^{n_1} \gamma_i^{d} \right)  \left| \Psi + (X \Gamma) (X \Gamma)^\top \right| ^{\frac{-\nu + n_1}{2}}.
\end{equation}

\subsection{Selection of the start temperature and cooling rate} \label{supp:tempering}

The start temperature $T_0$, and the cooling rate $t_c$, must be chosen carefully. If $T_0$ is too low , then the chain will not be able to explore the state space freely, and become trapped in a local minima early in the simulation. Alternatively, if $T_0$ is initiated too high, sampling is inefficient with too many samples obtained from the priors in early iterations of the tempering regime. 

To inform our selection of $T_0$ appropriately, we first sample randomly from the priors on $\mathbf{\theta}$ and $\mathbf{\gamma}$ and propose random $P$ vectors. We then calculate $\Phi$, the negative log of the posterior predictive, for each combination. We then calculate $T_0$ as
\begin{equation}
T_0 = \frac{P_{95}-P_{5}}{\log{(1+\tau)}},
\end{equation}
where $\tau=0.01$, a user defined tolerance to govern how high a start temperature should be set and $P_{95}$ and $P_{5}$ the $95^{th}$ and $5^{th}$ percentiles of the calculated $\Phi$ values. We then set the cooling rate to equal 
\begin{equation}
t_c=T_0^{-f_c/N}
\end{equation}
where $f_c=2000$ is the minimum number of iterations performed at one temperature, and $N$ is the original, user-defined number of iterations of the algorithm. We choose $f_c$ to ensure accurate calculation of acceptance rates between temperature changes and enable the system to equilibriate between each decrease in temperature. The temperature is decreased as
\begin{equation}
T'=T t_c, 
\end{equation}
using an exponential multiplicative cooling regime as proposed in~\cite{kirkpatrick1983}. 

The temperature is reduced every $f_c$ iterations until $T=1$, where we then perform $N_1$ samples at $T=1$, the assumed un-tempered posterior distribution. It is crucial that the reduction of $T$ is sufficiently slow, otherwise it is likely that the chain will become trapped in some local minima of the state space. We therefore impose the condition that $T$ is only decreased when the acceptance rate of proposals on the model transformation parameters is 23.4$\pm 10\%$ to ensure that we are sampling efficiently at any given instance of the tempered posterior, before progressing to a different temperature. 

\subsection{Random-walk proposals on bounded parameters} \label{supp:boundedRW}
To propose moves on bounded parameters, we transform the bounded parameters, $\mu_b$, to some transformed parameter $\mu_t$ using  
\begin{equation}
    \mu_t = T(\mu_b) = \log \left( \frac{\ell_\uparrow - \ell_\downarrow}{\mu_b - \ell_\downarrow} -1 \right),
\end{equation}
where $\ell_\uparrow$ and $\ell_\downarrow$ are the upper and lower limits of the bounded parameter respectively. After transformation to $\mu_t$, we can then perform standard random walk proposals on $\mu_t$ using equation~\ref{eq:rw}, where the corresponding prior variance is replaced by the variance on $\mu_t$, and use the inverse map $T^{-1}$ to map back to the new proposed $\mu_b$ 
\begin{equation}
    T^{-1} = \frac{\ell_\uparrow - \ell_\downarrow}{\exp(\mu_t) + 1} + \ell_\downarrow.
\end{equation}
The resulting proposal is no longer a random walk, and the proposals are no longer symmetric. The ratio of the proposal densities as required in the Metropolis-Hastings formula is given by:
\begin{equation}
    \frac{Q(u|v)}{Q(v|u)} = \prod_{i=1}^{n_p} \frac{e^v_i}{e^u_i} \left(\frac{e^u+1}{e^v+1} \right)^2,
\end{equation}
where $n_p$ is the number of parameters being sampled on. 

\subsection{Adaptive MCMC}\label{supp:adaptivemcmc}

When sampling at T=1, we switch to an adaptive sampling approach to increase efficiency of our sampling. The sample covariance is calculated on-the-fly to avoid the need for storing all accepted parameter values using 
\begin{equation} 
C_{i+1}= \left(\frac{i-2}{i-1} \right) C_i + \left(\frac{1}{i}\right) \Delta_\mathbf{\theta} \Delta_\mathbf{\theta} ^\top,
\end{equation} 
where $i$ is the iteration number during sampling at $T=1$, $\Delta_\mathbf{\theta} = (\mathbf{\theta} - \bar{\theta}$) and  $\bar{\theta}$ the rolling average of the model parameters. The proposal covariance matrix is updated less frequently as $i \rightarrow N_1$ to satisfy the diminishing adaptation condition. $C$ is then used as the proposal covariance matrix to generate future proposals.

\subsection{Estimation of transformation parameter and fidelity parameter MAP estimates conditioned on MLM} \label{supp:mapest}
During non-tempered sampling (at T=1) the minimum value of the negative log of the posterior density from equation~\ref{eq:finalposterior} is stored, along with corresponding $\theta$ and $\gamma$ parameter values. This gives us an estimate of the the deepest mode within the explored state space. To estimate the MAPs of model parameters conditioned on the MLM, we use the inbuilt \code{fmincon} optimiser in MATLAB, using starting positions of the parameters as those identified at the minimum log of the posterior density. The permutation vector is not changed from the MLM during optimisation as optimisation over the discrete permutation vector state space would have been computationally expensive and likely unnecessary due to the low acceptance of new permutation vectors during sampling at $T=1$. The maximum number of iterations and evaluations of the function were set to $10^6$. These values of $\theta$ and $\gamma$ are then used to generate spatial matching figures and displayed alongside permutation heatmaps.

\subsection{Implementation} \label{supp:systemspecs}
Matching of cells/landmarks was performed using MATLAB2020a on a system comprised of an Intel(R) Xeon(R) CPU E5-4627 v2 @ 3.30GHz (32 core), 62GB RAM. Post-analyses and figures were constructed using MATLAB2020a.

\subsection{Fidelity parameter test with higher $n_r$}\label{supp:fid20perc}
To further investigate the effect of the fidelity parameters (in addition to tests in Section~\ref{sec:untruematching}), we designed a more challenging test. We instead removed $n_r=6$ and $12$ cells from $Y^1$ and $Y^2$ for the 33- and 62-cell examples respectively. 

\begin{figure}[htpb]
    \centering 
    \includegraphics[width=\textwidth]{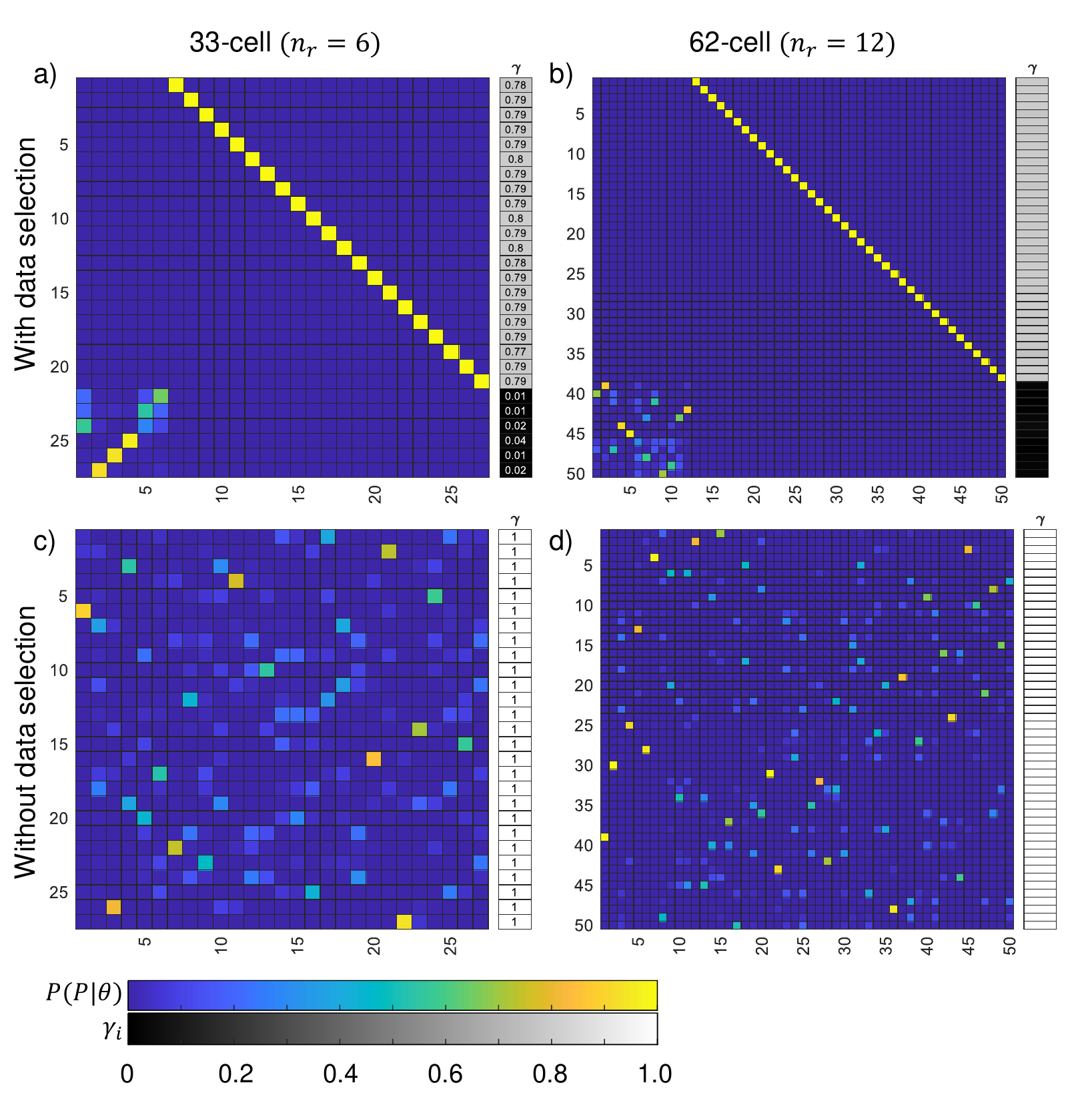}
    \caption{Example of permutation heatmaps for the test-problems included in Section~\ref{sec:untruematching} and~\ref{supp:fid20perc}, where $n_r=6,12$ cells were removed from the 33-cell and 62-cell datasets respectively. a-b) Example permutation heatmaps for the 32 and 62-cell problems when data selection is included. c-d) Example permutation heatmaps when data selection is not included for the two tests, MLMs found to have 21 and 37 incorrect matches for these particular examples.}
    \label{suppfig:fid20perc}
\end{figure}

When data selection was included, all chains sampled from distributions highly concentrated about the correct permutation vector for cells $1 \rightarrow (n_1-n_r)$. The final $n_r$ cells has reduced posterior means of the fidelity parameters and exhibited non-committal matching in the permutation probability heatmap,  Figure~\ref{suppfig:fid20perc}a-b. 

When we attempted to identify the matching of the points without data selection, matching success was reduced with increased numbers of incorrect matches in the MLMs, see  Figure~\ref{suppfig:fid20perc}c-d, and increased median cell-to-match distances, see tables~\ref{tab:fid_all_with} and~\ref{tab:fid_all_without}. 

\subsection{Embryo microinjection} \label{supp:microinj}
Embryo microinjection was performed using the protocols outlined in~\cite{alithesis2017}. For a more detailed description of methods please refer to the full materials and methods within~\cite{alithesis2017} and~\cite{Plusa2005}. Reference markers inside embryos were generated through microinjection of H2b-mCherry mRNA (1$\mu g/\mu l$) into single blastomeres of 8-cell stage embryos (collected from oviducts via flushing). Embryos were placed in drops of M2 in the deepest point of a glass-depression slide (Fisher), mounted within the micro-injector microscope stage (Leica). Drops of M2 were then covered with mineral oil to prevent evaporation. 

Embryos were held using blunted, pulled 1mm glass capillaries (Harvard apparatus GC100T-10) filled with M2. Cells were injected using non-blunted, pulled 1mm glass capillaries with internal filaments  (Harvard apparatus, GC100TF-10) using Sutter P97 needle puller. Needles were loaded with H2b-mCherry mRNA and selected cells injected using the Eppendorf FemtoJet injector unit. 

Embryos were then washed in KSOM and cultured in 35-mm glass bottom culture dishes for 24 hours at 37.5$^\circ$C and 5$\%$ CO$_2$.

\clearpage
\section*{Supplementary Tables}

\begin{table}[H] 
    \centering
    \caption{Parameters used to generate \textit{in silico} data sets.\\ }
\resizebox{\textwidth}{!}{%
\begin{tabular}{|p{2.5cm} | p{10.5cm}|}
\hline
Parameter & Value   \\ \hline
$A$          & $\begin{bmatrix} -0.9382  & -0.0621 &   0.0725 \\     0.0834 &  -0.1912  &  0.9195\\    -0.0361  &  0.8969  &  0.1846 \end{bmatrix}$ \\ \hline
$b$          & $\begin{bmatrix} 0.0178\\    -0.0197\\    0.0586\end{bmatrix}$                                                                     \\ \hline
$p_{x33}$ & [1.69,0,0,0,0.495,-1.31,0,0.315,-0.326,0.160,0,1.48,-0.945,-2.36,0.226,-1.01,-0.400,0,1.39,0,-1.75,0,0,0,-0.397,0.178,0,-1.64,0,0.741,0,0,0]                        \\ \hline
$p_{y33}$ & [0.365,0,0,0,0.664,-0.742,0,0.790,0.285,-2.12,0,0.699,-0.793,-1.66,0.218,1.22,-0.442,0,-0.619,0,-0.472,0,0,0,-1.79,-0.434,0,0.467,0,1.09,0,0,0]                          \\ \hline
$p_{z33}$ & [-1.10,0,0,0,-0.710,-1.47,0,-0.801,1.31,0.707,0,0.159,-2.05,-0.958,-0.823,0.156,0.448,0,-0.013,0,-0.060,0,0,0,-0.267,0.465,0,0.112,0,0.757,0,0,0]                         \\ \hline
\end{tabular}}

\label{tab:datagenparamvalues}
\end{table}

\begin{table}[htbp]
\centering
\caption{Summary of the number of incorrect matches in the MLM, median and RMSE cell-to-match distances ($d$) for each chain, for test problems in Section~\ref{sec:untruematching} when data selection was included. \\}
\resizebox{\textwidth}{!}{%
\begin{tabular}{ll|llllllll|}
\cline{3-10}
\multicolumn{2}{l|}{\textbf{WITH DATA SELECTION}}  & 1     & 2     & 3     & 4     & 5     & 6     & 7     & 8     \\ \hline
\multicolumn{1}{|l}{33-cell}  & \#incorrect matches  & 0     & 0     & 0     & 0     & 0     & 0     & 0     & 0     \\
\multicolumn{1}{|l}{$n_r=3$}  & median($d$)    & 0.029 & 0.029 & 0.029 & 0.030 & 0.028 & 0.030 & 0.035 & 0.029 \\
\multicolumn{1}{|l}{}         & RMSE($d$)       & 2.033 & 2.031 & 2.033 & 2.023 & 2.026 & 2.032 & 2.036 & 2.029 \\ \hline
\multicolumn{1}{|l}{62-cell}  & \#incorrect matches  & 0     & 0     & 0     & 0     & 0     & 0     & 0     & 0     \\
\multicolumn{1}{|l}{$n_r=6$}  & median($d$)     & 0.022 & 0.022 & 0.031 & 0.030 & 0.023 & 0.025 & 0.028 & 0.023 \\
\multicolumn{1}{|l}{}         & RMSE($d$)       & 1.807 & 1.804 & 1.807 & 1.807 & 1.811 & 1.810 & 1.810 & 1.803 \\ \hline
\multicolumn{1}{|l}{33-cell}  & \#incorrect matches  & 0     & 0     & 0     & 0     & 0     & 0     & 0     & 0     \\
\multicolumn{1}{|l}{$n_r=6$}  & median($d$)     & 0.046 & 0.047 & 0.079 & 0.044 & 0.046 & 0.044 & 0.044 & 0.049 \\
\multicolumn{1}{|l}{}         & RMSE($d$)       & 1.777 & 1.776 & 1.721 & 1.792 & 1.781 & 1.783 & 1.794 & 1.781 \\ \hline
\multicolumn{1}{|l}{62-cell}  & \#incorrect matches  & 0     & 0     & 0     & 0     & 0     & 0     & 0     & 0     \\
\multicolumn{1}{|l}{$n_r=12$} & median($d$)     & 0.028 & 0.041 & 0.028 & 0.033 & 0.028 & 0.028 & 0.035 & 0.032 \\
\multicolumn{1}{|l}{}         & RMSE($d$)       & 1.763 & 1.805 & 1.795 & 1.800 & 1.785 & 1.781 & 1.801 & 1.790 \\ \hline
\end{tabular}}

\label{tab:fid_all_with}
\end{table}

\begin{table}[htbp]
\centering
\caption{Summary of the number of incorrect matches in the MLM, median and RMSE cell-to-match distances ($d$)  for each chain, for each test problem in Section~\ref{sec:untruematching} when data selection was not included.\\}
\resizebox{\textwidth}{!}{%
\begin{tabular}{ll|llllllll|}
\cline{3-10}
\multicolumn{2}{l|}{\textbf{WITHOUT DATA SELECTION}} & 1     & 2     & 3     & 4     & 5     & 6     & 7     & 8     \\ \hline
\multicolumn{1}{|l}{33-cell}  & \#incorrect matches  & 2     & 2     & 2     & 2     & 2     & 2     & 2     & 2     \\
\multicolumn{1}{|l}{$n_r=3$}  & median($d$)    & 0.657 & 0.690 & 0.713 & 0.666 & 0.645 & 0.698 & 0.706 & 0.701 \\
\multicolumn{1}{|l}{}         & RMSE($d$)      & 1.772 & 1.771 & 1.722 & 1.776 & 1.794 & 1.741 & 1.760 & 1.765 \\ \hline
\multicolumn{1}{|l}{62-cell}  & \#incorrect matches   & 49    & 14    & 12    & 49    & 12    & 14    & 14    & 13    \\
\multicolumn{1}{|l}{$n_r=6$}  & median($d$)     & 1.317 & 0.892 & 0.683 & 1.317 & 0.693 & 0.685 & 0.700 & 0.896 \\
\multicolumn{1}{|l}{}         & RMSE($d$)       & 0.591 & 0.741 & 1.039 & 0.576 & 0.659 & 0.652 & 0.645 & 0.697 \\ \hline
\multicolumn{1}{|l}{33-cell}  & \#incorrect matches   & 15    & 3     & 7     & 19    & 21    & 19    & 7     & 20    \\
\multicolumn{1}{|l}{$n_r=6$}  & median($d$)    & 1.985 & 1.301 & 2.087 & 2.529 & 2.617 & 2.458 & 2.159 & 2.464 \\
\multicolumn{1}{|l}{}         & RMSE($d$)       & 1.298 & 1.263 & 1.119 & 1.369 & 1.586 & 1.389 & 1.154 & 1.365 \\ \hline
\multicolumn{1}{|l}{62-cell}  & \#incorrect matches  & 37    & 8     & 7     & 37    & 8     & 33    & 9     & 8     \\
\multicolumn{1}{|l}{$n_r=12$} & median($d$)     & 1.625 & 0.873 & 0.927 & 1.348 & 0.909 & 1.507 & 0.882 & 0.880 \\
\multicolumn{1}{|l}{}         & RMSE($d$)       & 0.916 & 0.944 & 0.925 & 0.882 & 0.930 & 0.842 & 0.840 & 0.933 \\ \hline
\end{tabular}}

\label{tab:fid_all_without}
\end{table}

\begin{table}[htbp]
\centering
\caption{Summary statistics for all chains from the non-linear deformation testing in Section~\ref{sec:deftests}. Here $-\log(\pi(\theta |Y^2, P_{\mathrm{MLM}}))$ 
is the negative log of the (un-normalised) posterior distribution post optimisation (conditioned on MLM), the number of incorrect matches is determined using the MLM, and $d_a$, $d_u$ correspond to the cell-to-match distances of all cells and the undeformed cells respectively. \\ }
\resizebox{\textwidth}{!}{%
\begin{tabular}{ll|cccccccc|}
\cline{3-10}
\textbf{Test}                     & \textbf{}             & \textbf{1} & \textbf{2} & \textbf{3} & \textbf{4} & \textbf{5} & \textbf{6} & \textbf{7} & \textbf{8} \\ \hline
\multicolumn{1}{|l|}{\textbf{a)}} & $-\log(\pi(\theta |Y^2, P_{\mathrm{MLM}}))$              & 192.791    & 192.791    & 192.791    & 192.791    & 192.791    & 192.791    & 192.791    & 192.791    \\
\multicolumn{1}{|l|}{\textbf{}}   & \#incorrect matches    & 2          & 2          & 2          & 2          & 2          & 2          & 2          & 2          \\
\multicolumn{1}{|l|}{\textbf{}}   & median($d_a$) & 1.303      & 1.294      & 1.304      & 1.289      & 1.290      & 1.305      & 1.302      & 1.309      \\
\multicolumn{1}{|l|}{\textbf{}}   & RMSE($d_a$)   & 1.070      & 1.050      & 1.064      & 1.081      & 1.079      & 1.060      & 1.093      & 1.076      \\
\multicolumn{1}{|l|}{\textbf{}}   & median($d_u$) & 0.538      & 0.522      & 0.531      & 0.524      & 0.516      & 0.529      & 0.533      & 0.530      \\
\multicolumn{1}{|l|}{\textbf{}}   & RMSE($d_u$)   & 1.364      & 1.347      & 1.356      & 1.379      & 1.379      & 1.355      & 1.390      & 1.376      \\ \hline
\multicolumn{1}{|l|}{\textbf{b)}} & $-\log(\pi(\theta |Y^2, P_{\mathrm{MLM}}))$                & 105.709    & -33.142    & 46.068     & -32.866    & 21.398     & 85.834     & 28.828     & -10.776    \\
\multicolumn{1}{|l|}{\textbf{}}   & \#incorrect matches    & 12         & 4          & 19         & 0          & 6          & 3          & 13         & 9          \\
\multicolumn{1}{|l|}{\textbf{}}   & median($d_a$) & 1.295      & 0.490      & 1.620      & 0.607      & 0.642      & 1.037      & 1.108      & 0.738      \\
\multicolumn{1}{|l|}{\textbf{}}   & RMSE($d_a$)   & 0.934      & 0.456      & 1.628      & 0.474      & 0.542      & 0.619      & 0.867      & 0.605      \\
\multicolumn{1}{|l|}{\textbf{}}   & median($d_u$) & 0.864      & 0.404      & 1.284      & 0.496      & 0.552      & 0.882      & 0.834      & 0.609      \\
\multicolumn{1}{|l|}{}            & RMSE($d_u$)   & 1.008      & 0.462      & 1.664      & 0.489      & 0.556      & 0.639      & 0.935      & 0.618      \\ \hline
\multicolumn{1}{|l|}{\textbf{c)}} & $-\log(\pi(\theta |Y^2, P_{\mathrm{MLM}}))$                & 93.737     & 74.685     & 180.749    & 9.301      & 74.136     & 38.764     & 21.887     & 211.862    \\
\multicolumn{1}{|l|}{}            & \#incorrect matches    & 6          & 7          & 32         & 2          & 6          & 2          & 2          & 28         \\
\multicolumn{1}{|l|}{}            & median($d_a$) & 0.373      & 0.613      & 2.113      & 0.206      & 0.415      & 0.335      & 0.306      & 1.845      \\
\multicolumn{1}{|l|}{}            & RMSE($d_a$)   & 0.571      & 1.254      & 2.298      & 0.462      & 1.200      & 0.855      & 0.979      & 2.015      \\
\multicolumn{1}{|l|}{}            & median($d_u$) & 0.371      & 0.487      & 2.515      & 0.190      & 0.346      & 0.300      & 0.275      & 2.523      \\
\multicolumn{1}{|l|}{}            & RMSE($d_u$)   & 0.573      & 1.317      & 2.353      & 0.474      & 1.248      & 0.881      & 1.010      & 2.051      \\ \hline
\multicolumn{1}{|l|}{\textbf{d)}} & $-\log(\pi(\theta |Y^2, P_{\mathrm{MLM}}))$               & 130.231    & 129.255    & 130.237    & 130.231    & 129.340    & 129.255    & 130.231    & 129.258    \\
\multicolumn{1}{|l|}{}            & \#incorrect matches    & 6          & 5          & 6          & 6          & 5          & 5          & 6          & 5          \\
\multicolumn{1}{|l|}{}            & median($d_a$) & 1.239      & 1.252      & 1.275      & 1.233      & 1.267      & 1.265      & 1.251      & 1.227      \\
\multicolumn{1}{|l|}{}            & RMSE($d_a$)   & 1.458      & 1.467      & 1.454      & 1.455      & 1.455      & 1.471      & 1.441      & 1.470      \\
\multicolumn{1}{|l|}{}            & median($d_u$) & 0.143      & 0.141      & 0.147      & 0.147      & 0.139      & 0.150      & 0.149      & 0.147      \\
\multicolumn{1}{|l|}{}            & RMSE($d_u$)   & 1.845      & 1.858      & 1.842      & 1.843      & 1.844      & 1.859      & 1.830      & 1.857      \\ \hline
\end{tabular} }
\label{tab:deftestsummary}
\end{table}

\clearpage
\section*{Supplementary Figures}

\end{document}